\documentclass[a4paper,11pt]{article}
\pdfoutput=1 

\usepackage{jcappub} 

\usepackage[T1]{fontenc} 
\usepackage{enumitem}
\usepackage{multirow}
\usepackage{tabularx}
\usepackage{rotating}
\usepackage[dvipsnames]{xcolor}
\usepackage{hhline}
\usepackage{afterpage}
\usepackage{placeins}
\usepackage{hyperref}
\usepackage{booktabs}
\usepackage{lipsum}
\usepackage{aasmacros}
\usepackage[normalem]{ulem}
\usepackage{float}
\usepackage{changepage}
\newcommand{\num}[2]{{#1}\mathrm{e}{#2}}
\usepackage{lineno}
\usepackage{caption}
\usepackage{subcaption}
\usepackage{placeins}
\usepackage{hyperref}
\usepackage[dvipsnames]{xcolor}

\graphicspath{{figures/}} 

\usepackage{natbib}
\usepackage[symbol]{footmisc}

\title{\boldmath Simulation-based inference of deep fields: galaxy population model and redshift distributions 
}

\renewcommand{\thefootnote}{\arabic{footnote}}

\author[a,1]{Beatrice~Moser,%
\note[1]{Corresponding author.}}
\author[a,b]{Tomasz~Kacprzak,}
\author[a]{Silvan~Fischbacher,}
\author[a]{Alexandre~Refregier,}
\author[a]{Dominic~Grimm,}
\author[c]{Luca~Tortorelli}

\vspace{0.2cm}

\affiliation[a]{Institute for Particle Physics and Astrophysics, ETH Zurich, Wolfgang-Pauli-Strasse 27, CH-8093 Zurich, Switzerland}
\affiliation[b]{Swiss Data Science Center, Paul Scherrer Institute, Forschungsstrasse 111, 5232 Villigen,
Switzerland}
\affiliation[c]{University Observatory, Faculty of Physics, Ludwig-Maximilian-Universit{\"a}t M{\"u}nchen,
Scheinerstrasse 1, 81679 Munich, Germany}

\emailAdd{moserb@phys.ethz.ch}
\emailAdd{tomaszk@phys.ethz.ch}
\emailAdd{silvanf@phys.ethz.ch}
\emailAdd{alexandre.refregier@phys.ethz.ch}
\emailAdd{dominic.grimm@phys.ethz.ch}
\emailAdd{ luca.tortorelli@physik.lmu.de}

\newcommand{\SExtractor}{\textsc{SExtractor}} 
\newcommand{\ufig}{\textsc{UFig}} 
 
\newcommand{\kcorrect}{\textsc{Kcorrect}} 

\abstract{Accurate redshift calibration is required to obtain unbiased cosmological information from large-scale galaxy surveys. 
In a forward modelling approach, the redshift distribution $n(z)$ of a galaxy sample is measured using a parametric galaxy population model constrained by observations. We use a 
model that captures the redshift evolution of the galaxy luminosity functions, colours, and morphology, for red and blue samples.
We constrain this model via simulation-based inference, using factorized Approximate Bayesian Computation (ABC) at the image level.
We apply this framework to HSC deep field images, complemented with photometric redshifts from COSMOS2020. 
The simulated telescope images include realistic observational and instrumental effects. 
By applying the same processing and selection to real data and simulations, we obtain a sample of $n(z)$ distributions from the ABC posterior. The photometric properties of the simulated galaxies are in good agreement with those from the real data, including magnitude, colour and redshift joint distributions. We compare the posterior $n(z)$ from our simulations to the COSMOS2020 redshift distributions obtained via template fitting photometric data spanning the wavelength range from UV to IR. We mitigate sample variance in COSMOS by applying a reweighting technique. We thus obtain a good agreement between the simulated and observed redshift distributions, with a difference in the mean at the 1$\sigma$ level up to a magnitude of 24 in the $i$ band. We discuss how our forward model can be applied to current and future surveys and be further extended. The ABC posterior and further material will be made publicly available at \href{https://cosmology.ethz.ch/research/software-lab/ufig.html}{this url}.}

\begin{document}
{
\renewcommand{\thefootnote}{\fnsymbol{footnote}}
\maketitle
\flushbottom
}

\newpage
\FloatBarrier
\section{Introduction}
Cosmological probes allow us to investigate the structure and components of our Universe, by posing constraints on a cosmological model. The standard model of cosmology, known as $\Lambda$CDM, comprises of three main components: dark energy, dark matter and baryons. These components can be traced in a large-scale galaxy survey, by measuring the positions and shapes of galaxies and their correlations. In recent years, state-of-the-art experiments such as the Dark Energy Survey\footnote{http://www.darkenergysurvey.org/} (DES; \cite{Abbot_2016}), the Kilo-Degree Survey\footnote{http://kids.strw.leidenuniv.nl/} (KiDS; \cite{de_Jong_2012}) and the Hyper Suprime-Cam Subaru Strategic Program \footnote{https://hsc.mtk.nao.ac.jp/ssp/survey/} (HSC; \cite{Aihara_2017}) have reported their constraints resulting from galaxy clustering, cosmic shear and galaxy-galaxy lensing and their combination, known as $3\times2$ point analysis \citep{Abbott_2022, Heymans_2021, More_2023}. Precise determination of the redshift distribution $n(z)$ of samples of galaxies is critical for obtaining cosmological constraints from galaxy surveys. Redshift information allows the separation of source and lens sample, and the computation of cosmological observables. Spectroscopy is prohibitively time-consuming as a technique to provide accurate redshifts of all galaxies in a wide survey, and is furthermore subject to selection biases, especially for faint samples. Surveys thus rely on integrated measurements in a limited number of broad-bands in order to determine the redshift distribution of the sample of interest (for review, see \cite{Newman_2022,Salvato_2019}). This has proven to be a challenging task, especially since the relationship between redshift and colour in a limited wavelength range is subject to degeneracies \cite{Salvato_2019}. The characterization of photometric redshift (photo-$z$) distributions is one of the key systematics affecting cosmic shear measurements since errors in the calibration of redshift distributions and their uncertainties can lead to biases in the retrieved cosmological parameters  \cite{Huterer_2006,Cunha_2012,Huterer_2013,Joudaki_2016, Hoyle_2018,Salvato_2019,Joudaki_2020, Fischbacher_2022}. Traditional photo-$z$ approaches  include template fitting (for example LePhare \citep{Arnouts_1999, Ilbert_2006}, BPZ \citep{Benitez_2000}, ZEBRA \citep{Feldmann_2006} and \textsc{EAZY} \citep{Brammer_2008})  and machine learning methods (for example ANNz \citep{Collister_2004}, ANNz2 \citep{Carrasco_Kind_2014, Sadeh_2016} and DNF \citep{De_Vicente_2016}). 

In a cosmological survey it is common to employ methods to constrain the overall redshift distribution of the sample of interest rather than the redshifts of single objects, either by an empirical reweighting of a well measured redshift sample \citep{Lima_2008, Hildebrandt_2020, Wright_2020}, by using spatial cross-correlations \citep{Gatti_2022}, or by a combination of the two \citep{, Myles_2021, Rau_2022}. Another approach, which has been introduced in recent years, is simulation-based inference (SBI) \citep{Herbel_2017, Kacprzak_2020, Alsing_2022}. SBI relies on forward modelling the survey of interest: the redshift distribution of a sample of galaxies is the result of the statistical properties of the observed galaxy population, the observing conditions and limitations of the detector, and the selection applied to define the target sample.  Accurate modelling of magnitude, colour and (at second-order) size distributions of galaxies as a function of redshift, taking into account the observational and instrumental effects, enables robust determination of the redshift distribution of the sample, as well as a straightforward estimation of its uncertainty. Furthermore, the constraints on the galaxy population model provide some insights on the statistical nature of galaxies and allow a robust treatment of complicated selection functions. The methodology used in this paper has been developed in \cite{Herbel_2017} and further extended and applied in \cite{Kacprzak_2020} and \cite{Tortorelli_2020, Tortorelli_2021}. We refer to this forward modelling framework as Monte Carlo Control Loops (MCCL) \cite{Refregier_2014}. The method relies on an empirical galaxy population model to describe the intrinsic properties of galaxies and stars. We render the objects with photon shooting methods using an image simulator called Ultra Fast Image Generator (\ufig\ \citep{Berge_2013}) in a set of broad-bands described by the filter throughputs of the telescope used. Moreover, we simulate observational and instrumental effects such as sky and detector noise, point spread functions (PSF), reddening and saturation. In this way, we can post-process the simulations and the real images in the same way. We run  \SExtractor\ \citep{Bertin_1996} on both to obtain catalogs of objects and apply the same selection functions to both the simulated and the real catalog, which simplifies the treatment of selection biases. The model parameters are constrained using the observed data via Approximate Bayesian Computation (ABC), using distance measures that ensure that the photometric properties of the objects in simulations statistically agree with real data. The method has been used to simulate and perform a cosmic shear measurement of the Dark Energy Survey Year 1 \citep{Bruderer_2016,Bruderer_2018, Kacprzak_2020}, for redshift calibration on Subaru data \citep{Herbel_2017},  and to obtain the luminosity functions of blue and red galaxies at different redshifts with Canada-France-Hawaii Telescope Legacy Survey (CFHTLS \citep{Cuillandre_2012}) data \citep{Tortorelli_2020}. Furhermore, it has been applied to simulate the narrow band imaging of Physics of the Accelerating Universe survey (PAUS \citep{Marti_2014}) \citep{Tortorelli_2018, Tortorelli_2021} and galaxy spectra from the Sloan Digital Sky Survey (SDSS \citep{Blanton_2017}) CMASS sample \citep{Fagioli_2018, Fagioli_2020}.  

In this work, we use HSC Deep/UltraDeep (DUD) data \cite{Aihara_2022} and accurate many-band photometric redshifts from COSMOS2020 \citep{Weaver_2022} in order to obtain tight constraints on the model parameters at high redshift. Previous constraints to the model parameters extended to a redshift of $z \sim 1$, whereas in this work we explore the regime of Stage IV surveys. \cite{Sudek_2022} suggests that HSC data is the most powerful for constraining the Schechter parameters of the luminosity function and thus the redshift distribution of galaxies, because of its exquisite depth. 
As done in previous work, we use an ABC framework, with several important practical improvements.
After tuning the model, we use the obtained posterior parameters to simulate HSC DUD data in the COSMOS field and validate the $n(z)$ for different magnitude cuts. While the HSC DUD fields are deeper than current and upcoming wide-field surveys, the magnitude cuts that we consider in this work are consistent with the weak lensing source samples of current ( $i$ band magnitude cut between 23 and 24) and upcoming surveys (expected $i$ band magnitude cut between 24.5 and 25.5). Using our ABC posterior directly to calibrate the redshift distributions of these survey would entail running image simulations of the survey of interest using the parameters from the ABC posterior obtained in this work. The resulting $n(z)$ distribution would include errors propagated from the ABC posterior constraints and due to selection effects. Calibrating the model with deeper data has the advantage that objects above the magnitude cut are constrained, which is necessary to include noise and source confusion, thus making the sample selection realistic.

The paper is structured as follows. Section \ref{sec:data} describes the HSC DUD data and the COSMOS2020 catalog used both to tune the model and validate our results. In Section \ref{sec:method}, we describe the methodology and introduce the changes compared to previous work. Section \ref{sec:results} reports the results of our analysis. We conclude the paper in Section \ref{sec:conclusion}. We assume a standard $\Lambda$CDM cosmology with h=0.7, $\Omega_m$=0.3, and $\Omega_\Lambda$=0.7 throughout the paper.

\section{Data}
\label{sec:data}
In this section, we present the data used to constrain our model of the galaxy population and to validate the obtained redshift distributions. We rely on data from the Deep and Ultradeep layers of the third data release (PDR3) of the Hyper Suprime-Cam Subaru Strategic Program  (HSC) \cite{Aihara_2022}. In order to provide our model with additional redshift information and validate the $n(z)$, we complement the HSC data with accurate photo-$z$ estimates from the COSMOS2020 panchromatic photometric catalog \cite{Weaver_2022}. 
\subsection{Deep/UltraDeep data from HSC PDR3}
\label{sec:hsc_data}
HSC is a large multi-band imaging survey conducted with the $8.2$-metre Subaru telescope. It comprises of three layers: Wide, Deep and UltraDeep. The Wide layer covers 1470 $\mathrm{deg}^2$, considering partially observed areas in five broad-band filters ($g$,$r$,$i$,$z$,$y$). The Deep/UltraDeep (DUD) layers cover $\sim36$ $\mathrm{deg}^2$ in the five broad-band filters and four additional narrow-band filters. In this work, we use the publicly available coadded broad-band DUD images with local sky subtraction from PDR3\footnote{https://hsc-release.mtk.nao.ac.jp/doc/index.php/available-data\_\_pdr3/}. There are four different fields: COSMOS, DEEP2-3, SXDS+XMM-LSS and ELAIS-N1. Each field is separated in tracts which are equi-area rectangular regions on the sky, divided in 9$\times$9 patches comprising of 12~arcmin per side corresponding to 4200~pixels and overlapping by 100~pixels. We use all the available DUD patches for tuning the model. The dataset consists initially of roughly 1500~patches. We blacklist patches where:
\begin{itemize}
    \item more than 30\% of the image area is flagged as \texttt{NO\_DATA},
    \item more than 50\% of the image area is covered by the \texttt{BRIGHT\_OBJECT} mask,
    \item the image overlaps for more than 30\% of the area with another patch.
\end{itemize}
\texttt{NO\_DATA} and \texttt{BRIGHT\_OBJECT} masks correspond to flags 8 and 9 in the mask layer of the data. The overlap, on the other hand, is computed using the footprint of the images and by masking the pixels on the top or upper edge of each coadd that are also covered by another patch. After blacklisting, we retain a total of 746~patches.
\subsection{COSMOS2020 catalog}
\label{sec:cosmos_data}
The COSMOS2020 catalog \cite{Weaver_2022} consists of nearly $1$~million high quality photometric redshifts derived via template fitting of many broad and narrow band observations ranging from UV to IR wavelengths. There are four different publicly available catalogs, which differ in the method used for extracting photometry (\SExtractor, used in \textsc{Classic}, and \textsc{The Farmer}) and for the photometric redshift template fitting code (\texttt{LePhare} \citep{Arnouts_1999,Ilbert_2006} and \textsc{EAZY} \citep{Brammer_2008}). Since we use \SExtractor\ for the photometric measurement on the HSC data in our pipeline, we also work with the COSMOS2020 \textsc{Classic} catalog. We use both \texttt{LePhare} and \textsc{EAZY} photo-$z$s. We remove areas where the photometry is unreliable or with partial coverage by means of the \texttt{FLAG\_COMBINED} parameter thus reducing the area to $1.27 deg^2$. We select objects that have \texttt{MAG\_AUTO} < 99, \texttt{LePhare} (\texttt{lp\_zBEST}) or \textsc{EAZY} (\texttt{ez\_z\_phot}) photo-$z$  between 0 and 8 (removing \texttt{Nan} values), \texttt{LePhare} object type galaxy (\texttt{lp\_type=0}) and \SExtractor\ \texttt{FLAGS} < 4. 
The COSMOS2020 catalog is used both for providing redshift information while constraining the model and for validation. The validation sample is explained in the following section, while the reweighting procedure used during the ABC analysis is detailed in Section~\ref{sec:zsample}.

\subsection{Validation sample}
\label{sec:validation}
In order to build our validation sample, we compare the COSMOS2020 \texttt{COMBINED} footprint with the HSC DUD data in the COSMOS field and find 63 overlapping patches, out of which 56 are almost fully covered. 
We perform \SExtractor\ forced photometry on these coadds using the $i$ band for detection and match the obtained catalog with objects in the COSMOS2020 catalog by position and magnitude (using the magnitude \texttt{MAG\_APER} measured in a circular $3''$ diameter apeture). The \texttt{BRIGHT\_OBJECT} masks from HSC PDR3 are very conservative and cause a loss of roughly one third of the COSMOS2020 objects, since COSMOS2020 uses the less conservative HSC PDR2 masks. We compare the simulated final redshift distributions in the COSMOS field to both \texttt{LePhare} and \textsc{EAZY} photo-$z$s at three different magnitude cuts and with the selection defined in Appendix~\ref{sec:appendix_abc_run}. We describe our procedure to account for sample variance in COSMOS in the next subsection.

\subsubsection{Sample variance in COSMOS}
\label{sec:cosmos_sample_variance}
The COSMOS field only spans 2 $\deg^2$ of the sky. This means that, while the volume spanned by COSMOS observations is large due to the considerable depth \cite{Scoville_2007}, there are notable sample variance effects at low redshifts. In order to estimate the impact of sample variance, we assign COSMOS2020 photo-$z$s to all galaxies in the other deep fields as described in Section~\ref{sec:zsample} and look at the offset in mean redshift. This ensures that we span a larger area and the difference in depth is negligible when only considering galaxies with $i$ band magnitude below 25\footnote{we use the $i$ band aperture magnitude in a $3''$ circular aperture \texttt{MAG\_APER3\_i}, as described in Section~\ref{sec:sextraction}}. In order to also measure the scatter due to sample variance, we produce 10 sub-fields the size of COSMOS (56 images) and measure the standard deviation of the 10 mean redshifts. We prefer this approach to a standard Bootstrap in order to preserve the locality of the effect, which is due to the inhomogeneity of large-scale structure on small scales. The redshift offsets are reported in Table~\ref{tab:zshifts}.
We observe a mean redshift offset between the COSMOS field and the other deep fields of $\Delta z=\ \langle z_{\mathrm{COSMOS}} \rangle - \langle z_{\mathrm{deep}} \rangle\ \approx 0.015$ when we cut at $i$ band magnitude of 23, meaning that sample variance causes a bias for the brightest sample.

\subsubsection{Offset between \textsc{EAZY} and \texttt{LePhare}}
\label{sec:eazy_vs_lephare}
We notice that, when applying a simple magnitude cut in the $i$ band and the selection described in Appendix~\ref{sec:appendix_abc_run}, there is a systematic offset between \textsc{EAZY} and \texttt{LePhare} photometric redshifts from COSMOS2020. Similarly to sample variance, this effect has a stronger impact on the brightest sample where \textsc{EAZY} predicts systematically higher redshifts than \texttt{LePhare}. The systematic offset is $\Delta z = \langle z_{\mathrm{EAZY}} \rangle - \langle z_{\texttt{LePhare}} \rangle = 0.014$ for a magnitude cut at \texttt{MAG\_APER3\_i}=23 and 0.018 for a magnitude cut at \texttt{MAG\_APER3\_i}=24. The offset is negligible when cutting at \texttt{MAG\_APER3\_i}=25. We report these offsets in Table~\ref{tab:zshifts}.

\begin{table}[]
    \centering
    \begin{tabular}{|c|c|c|c|}
    \hline
    Value & $i$ band mag cut 23 &$i$ band mag cut 24&$i$ band mag cut 25\\
    \hline
    $\Delta z$ sample variance \texttt{LePhare} & $0.014 \pm 0.012$ & $0.001 \pm 0.019$ & $0.002 \pm 0.019$ \\
    $\Delta z$ sample variance \textsc{EAZY} & $0.015 \pm 0.012$ & $0.001 \pm 0.019$ & $0.003 \pm 0.018$\\
    $\Delta z$=$ \langle z_{\mathrm{EAZY}} \rangle - \langle z_{\texttt{LePhare}} \rangle$     & 0.014 & 0.018 & 0.0006 \\
    \hline
    \end{tabular}
    \caption{Shifts in photometric redshift due to sample variance and difference between the two photo-$z$ codes used in COSMOS2020.}
    \label{tab:zshifts}
\end{table}

\section{Method}
\label{sec:method}
The backbone of our forward modelling framework is an empirical parametric model of the galaxy population, used to generate distributions of intrinsic properties of galaxies. Once a galaxy catalog is generated given a set of model parameters, we simulate an image of the survey of interest, in our case the HSC DUD fields. In order to obtain a realistic simulation, we include the effects of the instrument and the known observational systematics that impact the photometric measurement. The end-to-end process from a set of model parameters to a realistic telescope image is implemented in the Ultra Fast Image Generator (\ufig\ \citep{Berge_2013}). \ufig\ has been developed as a simulator for MCCL, with speed as one of the primary features. Computational speed is critical for this task, since a large number of simulations is required to tune the parameters of the model. The inference is performed by running an Approximate Bayesian Computation (ABC), where the realism of our simulated images is increased by minimizing a set of distance measures. Despite the optimized simulator, the computational cost of performing a full ABC inference amounts to roughly a million CPU hours on the Euler HPC cluster of ETH Zurich. In the following, we describe the galaxy population model, how we extend \ufig\ to reproduce realistic HSC DUD images, how we include redshift information from COSMOS2020 reducing the impact of cosmic variance and the details of our ABC scheme. We focus on the novelties introduced in this work.
\subsection{Galaxy population model}
\label{sec:abc_updates_model}
\begin{figure}
    \centering
    \includegraphics[width=1\linewidth]{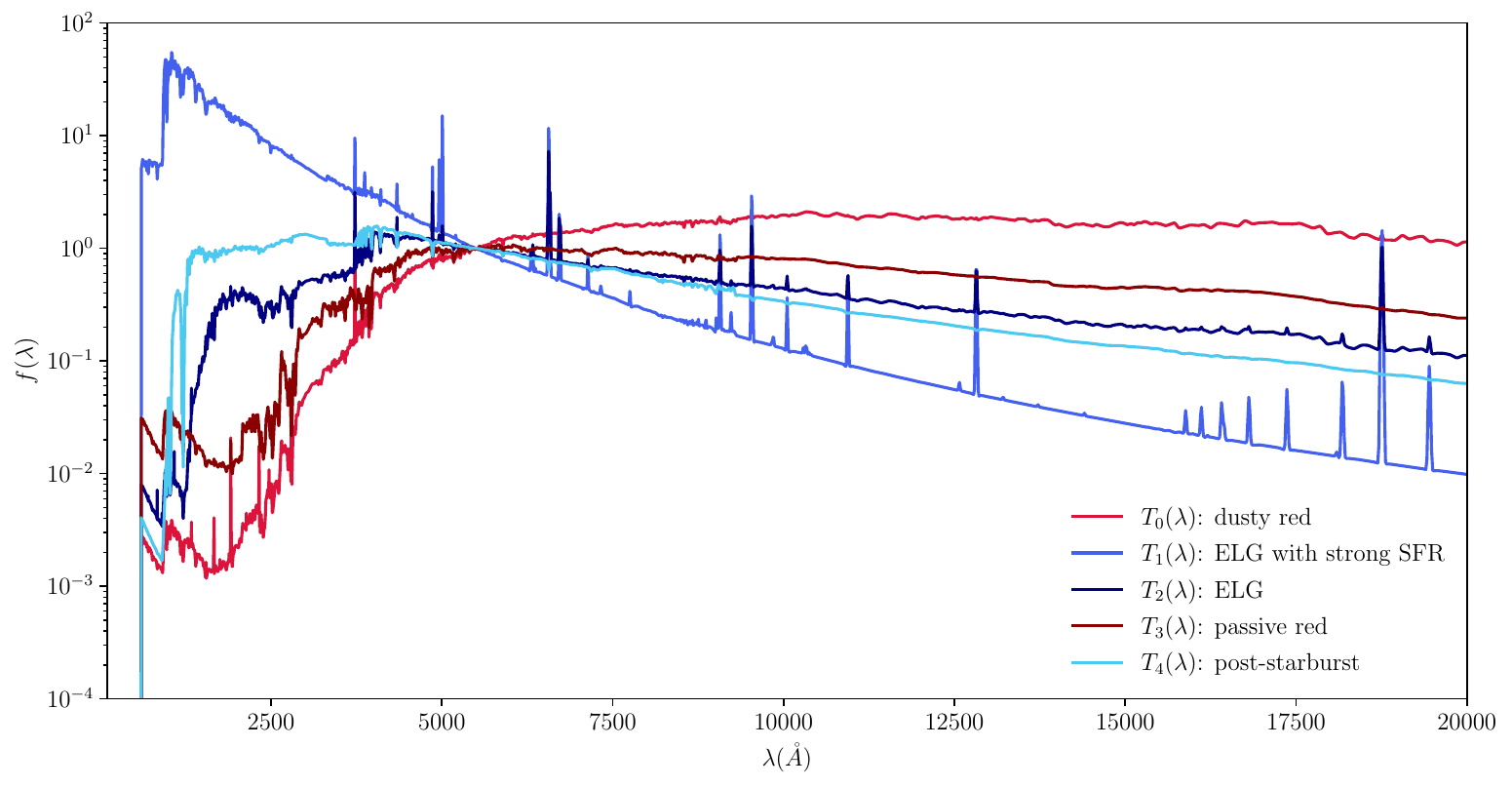}
    \caption{The five \kcorrect\ templates \cite{Blanton_2005} which are combined linearly to obtain the spectral energy distributions of galaxies. They are renormalized so that $f(\lambda)=1$ at $\lambda=5500 (\mathring{A})$.}
    \label{fig:kcorrect}
\end{figure}

We include in our model two different populations of galaxies: red and blue, often referred to as quiescent and star-forming galaxies. The separation between red and blue galaxies is intrinsic: we have separate sets of parameters in our model for the two galaxy populations. We sample absolute magnitudes $M$ and redshifts $z$ from Schechter luminosity functions 
\begin{equation}
\label{eq:schechter}
    \phi(z,M) = \frac{2}{5}{\ln{10}} \phi^*(z)10^{\frac{2}{5}(M^*(z)-M)(\alpha+1)}\exp{(-10^{\frac{2}{5}(M^*(z)-M)})},
\end{equation}
where $M^*(z)$ and $\phi^*(z)$ are functions of redshift and have separate parameters for blue and red galaxies.
 We then assign a spectral energy distribution (SED) to each galaxy as a linear combination of 5 spectral templates from \kcorrect\ \citep{Blanton_2017} $$SED(\lambda) = \sum_{i=0}^4c_iT_i(\lambda).$$ The coefficients of the templates are also different for blue and red galaxies and evolve with redshift. We show the five \kcorrect\ template spectra for reference in Figure~\ref{fig:kcorrect}. We assign sizes to galaxies using a log-normal distribution for the half light radius and a Sersic light profile. The ellipticities are sampled from a Beta distribution.  As explained in the following subsections, in this work all galaxy population parameters are constrained separately for red and blue galaxies. Furthermore, we add stars to our simulations using the Besan\c{c}on model of the Milky Way \citep{Robin_2004}. The magnitudes from the catalog of pseudo-stars are sampled with replacement, and the positions are assigned randomly within the \textsc{Healpix} pixel (\texttt{nside}=8, see \cite{Herbel_2017,Kacprzak_2020} for more details).
The positions of the bright end of the star distribution is taken from the Gaia DR3 catalog \citep{Gaia_2016, Gaia_2022} and abundance-matched to the Besan\c{c}on model. For an extensive description of the galaxy population model, see \citep{Herbel_2017, Tortorelli_2020}.

In the following, we highlight the modifications to the galaxy population model compared to \citep{Herbel_2017,Tortorelli_2020,Kacprzak_2020}:
(i) modification of the luminosity function parametrization, 
(ii) addition of new parameters in the morphology sector to allow different characteristics for blue and red galaxies,
(iii) small changes to the parametrization of ellipticities and Sersic indices,
(iv) modification of parametrization for template coefficients of the SED.

\subsubsection{Luminosity function parametrization}
We modify the redshift evolution of the luminosity function parameters, in accordance with galaxy evolution models \cite{Johnston_2011}. In the Pure Luminosity Evolution (PLE) scenario, massive galaxies assemble and form most of their stars at high redshifts. They then evolve without merging. This results in the functional form
\begin{equation}
    M^*(z) = M^*_\mathrm{intcpt} + M^*_\mathrm{slope}\log(1+z)
    \label{eq:m_star_evol}
\end{equation}
for the evolution of the characteristic absolute magnitude with different parameters $M^*_\mathrm{intcpt}$ and $M^*_\mathrm{slope}$ for blue and red galaxies. In the Pure Density Evolution (PDE) scenario, galaxies undergo mergers so that they are more massive but less numerous at lower redshifts. This scenario can be modelled through the evolution of the normalization of the luminosity function as
\begin{equation}
    \phi^*(z) = \phi^*_{\mathrm{ampl}}(1+z)^{\phi^*_{\mathrm{exp}}}
    \label{eq:phi_star_evol}
\end{equation}
where $\phi^*_{\mathrm{ampl}}$ and $\phi^*_{\mathrm{exp}}$ also depend on the galaxy population. We vary these 8 parameters during the ABC. Furthermore, $\alpha_{\mathrm{blue}}$ and $\alpha_{\mathrm{red}}$, describing the steepness of the faint-end slope of the luminosity function, are also varied in this analysis, differently from previous work. 
\subsubsection{Updated galaxy morphology}
We added new parameters in the morphology section.
The relation between galaxy half light radius $r_{50}$ and absolute magnitude is described by three parameters: $\log r_{50}^{\rm{intcpt}}$, $\log r_{50}^{\rm{slope}}$ and $\log r_{50}^{\rm{std}}$ \cite{Herbel_2017}. We sample  $\log r_{50}$ from a normal distribution with mean $$\log r_{50}^{\rm{mean}} = \log r_{50}^{\rm{slope}} M + \log r_{50}^{\rm{intcpt}}$$ and standard deviation $\log r_{50}^{\rm{std}}$, where $M$ is the absolute magnitude of the galaxy.  The distribution of $r_{50}$ is thus log-normal (as found in e.g. \citep{de_Jong_2000,Shen_2003}).
In our updated model we have a separate set of these parameters for red and blue galaxies.
We also constrain the $\log r_{50}^{\rm{std}}$ parameter during the ABC inference, which was fixed in previous work.

\subsubsection{Ellipticity and Sersic indices}
The parameterization of ellipticity $p(e)$ using a Beta distribution is slightly modified; parameters $e_{\rm{mode}}$, $e_{\rm{spread}}$ correspond to the mode and concentration of the Beta distribution respectively. They are related to the parameters of the Beta distribution $\alpha$ and $\beta$ through $e_{\rm{spread}}=\alpha+\beta$ and $e_{\rm{mode}}= \frac{\alpha -1}{e_{\rm{spread}}-2}$.
This change makes the parameters easier to interpret and allows for designing simpler priors.
We also modified the prescription for modelling the distribution of Sersic indices.
We use a Betaprime distribution with free parameter $n_{s}$, which is the mode of the  distribution.
It is related to parameters $\alpha$ and $\beta$ of the Betaprime distribution as $n_s = \frac{\alpha-1}{\beta+1}$.
Parameter $\beta$ is responsible for the scatter, and fixed throughout the analysis, to the following values:
$\beta_{\rm{blue}}=5$ and 
$\beta_{\rm{red}}=50$. These values were chosen so that the distributions of Sersic indices match that of \cite{Tarsitano_2018}.

\subsubsection{Spectral templates parametrization}
In our model, the coefficients of the SED templates are drawn from a five-dimensional Dirichlet distribution separately for blue and red galaxies, similarly to \citep{Herbel_2017, Kacprzak_2020, Tortorelli_2020}. The Dirichlet distribution is used because the samples drawn from it sum to 1 and the spectrum can then be rescaled to match the absolute magnitude of the galaxy. The parameters of the Dirichlet distribution evolve with redshift: we use two separate sets of parameters for
z=0 and z=3, with parameters for other redshifts being an interpolation between them $$ \alpha_i(z) = \left(\alpha_{i,0}\right)^{1-\frac{z}{3}} \times \left(\alpha_{i,3}\right)^{\frac{z}{3}}.$$ 
In previous work the $\alpha_i$ were constrained at redshifts $z=0$ and $z=1$; we now use $z=3$ since the functional form is fixed (so that we do not need to have a large sample of galaxies at $z=3$ to pose limits on the parameters' values) and this allows us to enforce prior bounds at higher redshifts.

Previously, the prior on this distribution was also a Dirichlet variable with unity weights, multiplied by a uniform number between [5,15], which accounted for the variance. This way, the $\alpha_{i}$ parameters were affecting both the mean and variance of the Dirichlet variable. We change the model to capture the mode and variance in separate parameters. Furthermore, \cite{Herbel_2017} derived weights to apply to each template using the New York University Value-Added Galaxy Catalog \citep{Blanton_2005} thus effectively using different template spectra for blue and red galaxies.
We removed the weights and reparametrized the template spectra to be normalized to 1 at wavelength $5500$ \AA.
We use a redundant parametrization with modes of the Dirichlet distributions $\bar \alpha_i, \ i=0, \dots, 4$ and two new $\alpha_{\rm{std,0/3}}$ parameters. The parameters $\alpha_{\rm{std,0/3}}$ correspond to the standard deviation of the 5-dimensional Dirichlet coefficients with equal concentrations at redshifts $z=0$ and $z=3$ and evolve in redshift the same way as the template coefficients $\alpha_i$.
We enforce the normalization $\sum_i \bar \alpha_i = 1$, and the final Dirichlet coefficients are calculated as $$\alpha_i =  1 + \bar \alpha_i \cdot \left[\frac{1}{N}\left(1-\frac{1}{N}\right)\alpha^{-2}_{\rm{std}} - N  - 1\right]$$ with $N$=5.
This new parametrization reduces the number of local minima in the problem and makes the $\bar \alpha_i$ variables more interpretable. Finally, the template spectra are the same for red and blue galaxies and we encapsulate the information about the different galaxy populations in the ABC prior on the template coefficients. The ABC prior on template coefficients  was obtained in a preprocessing step where we performed an ABC on catalog level using the COSMOS2015 catalog \cite{Laigle_2016}, as described in Appendix \ref{sec:appendix_abc_model}.
 The final model has 46 parameters, out of which 4 are redundant (see Table \ref{tab:abc_model} in Appendix \ref{sec:appendix_abc_model}).

\subsection{Image simulations of HSC DUD fields}
\label{sec:HSC-forward-model}

The catalogs of intrinsic galaxy properties are used to create simulated HSC DUD images. The image generation procedure, including realistic observational and instrumental effects, is as follows.
We input the metadata provided by the HSC database\footnote{https://hsc-release.mtk.nao.ac.jp/datasearch/} about size of the image in pixels, pixel scale ($0.168''/$pixel) and sky coordinates of the images. We perform our simulations using the $g$, $r$, $i$, $z$ and $y$ broad-bands. HSC replaced the $r$ and $i$ filters with more uniform filters $r2$ and $i2$ which have been coadded together with $r$ and $i$.  In our simulations, we use the filter throughputs from $r2$ and $i2$, after checking that the magnitude shifts are small. In order to compute the apparent magnitude of a galaxy in a specific broad-band, we integrate over its SED and the filter throughput taking into account k-corrections and reddening due to galactic extinction. The computation of arbitrary magnitudes in the AB system is described in Section 3.2.3 of \cite{Herbel_2017} and the wavelength dependent extinction to account for reddening in Appendix D of the same paper. The magnitude zeropoint is set to 27 mag/ADU for the HSC coadds.

We simulate PDR3 coadded images directly, as introduced in Section~\ref{sec:hsc_data}. 
In order to simulate the coaddition process, we use systematic maps derived from the metadata. We create a map of the exposure times and number of exposures per pixel for each patch. The CCD gain of a single exposure multiplied by the number of exposures per pixel gives us a rough estimate of the effective gain to convert between ADUs and number of photons. Galaxies are randomly distributed on the image and rendered by sampling individual photons according to the galaxy's Sersic profile. This procedure  naturally includes Poisson noise \cite{Berge_2013, Bruderer_2016}. The Point Spread Function (PSF) is rendered as a distortion to the light profile of the galaxy. In order to estimate the impact of the PSF in the real images, we use a Convolutional Neural Network (CNN) as presented in \cite{Herbel_2018} and updated in \cite{Kacprzak_2020}. The PSF is estimated at the position of stars matched with Gaia DR3 \citep{Gaia_2016, Gaia_2022} with magnitudes included between 18 and 22 in the $i$ band, which have \SExtractor\ \texttt{FLAGS} 0 or 16 and are not at the image boundary. We perform this selection because stars with apparent magnitudes lower than 18 in the $i$ band are included in the bright objects masks of HSC. The matching is done with a Balltree with a maximum distance of 1.5 pixels.  We reserve 15\% of the selected stars for validation. Each PSF parameter is then interpolated across the coadd using a Chebyshev polynomial basis of maximum order 4 (see Appendix C of \cite{Kacprzak_2020} for details). 
\begin{figure}
    \centering
    \includegraphics[width=1\linewidth]{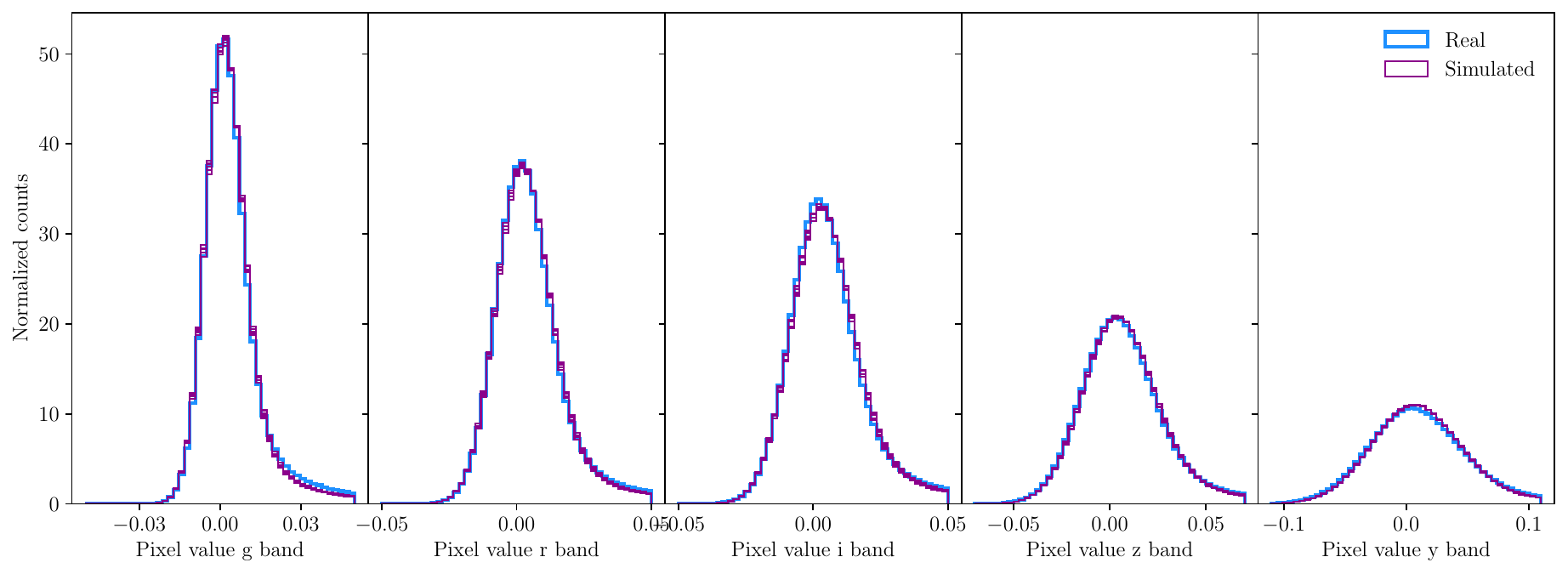}
    \caption{Example of pixel histogram around the background values from \texttt{tract} 9813 patch 702. The simulated background is shown from 5 different ABC configurations.}
    \label{fig:bkg}
\end{figure}

In order to simulate the background noise in an image, we first derive a map of the root-mean-square of the noise from the real data using the \texttt{Background2D} function with \texttt{SExtractorBackground} estimator and 3$\sigma$ sigma-clipping from \texttt{photutils} \citep{Bradley_2022}. This map is derived from the data individually for each patch. We then add Gaussian noise to the simulation with mean read off from the image header and standard deviation taken from the map (different for each pixel). Since the standard deviation of the noise that we apply is already different in each pixel and is estimated from background subtracted images, we do not need to perform any background subtraction, including local background subtraction which has the most impact in the surroundings of bright objects, which are anyway masked. The resulting simulated background is in good agreement with the background in the real data.
We show an example of the pixel histogram of an image for real data and simulations for low pixel values in Figure \ref{fig:bkg}. We observe that there is a slight overestimation of the background level due to the lack of background subtraction.

An alternative approach to background estimation would amount to adding a Gaussian background using the parameters in the image headers (both mean and standard deviation) and then applying global and local background subtraction. In our tests this  procedure worsened the agreement between data and simulations. 
We create masks of the areas with no data and surrounding bright stars using the bit flags 8 and 9 from the second layer of the fits files. 

\begin{figure}[h]
    \centering
    \includegraphics[width=1\linewidth]{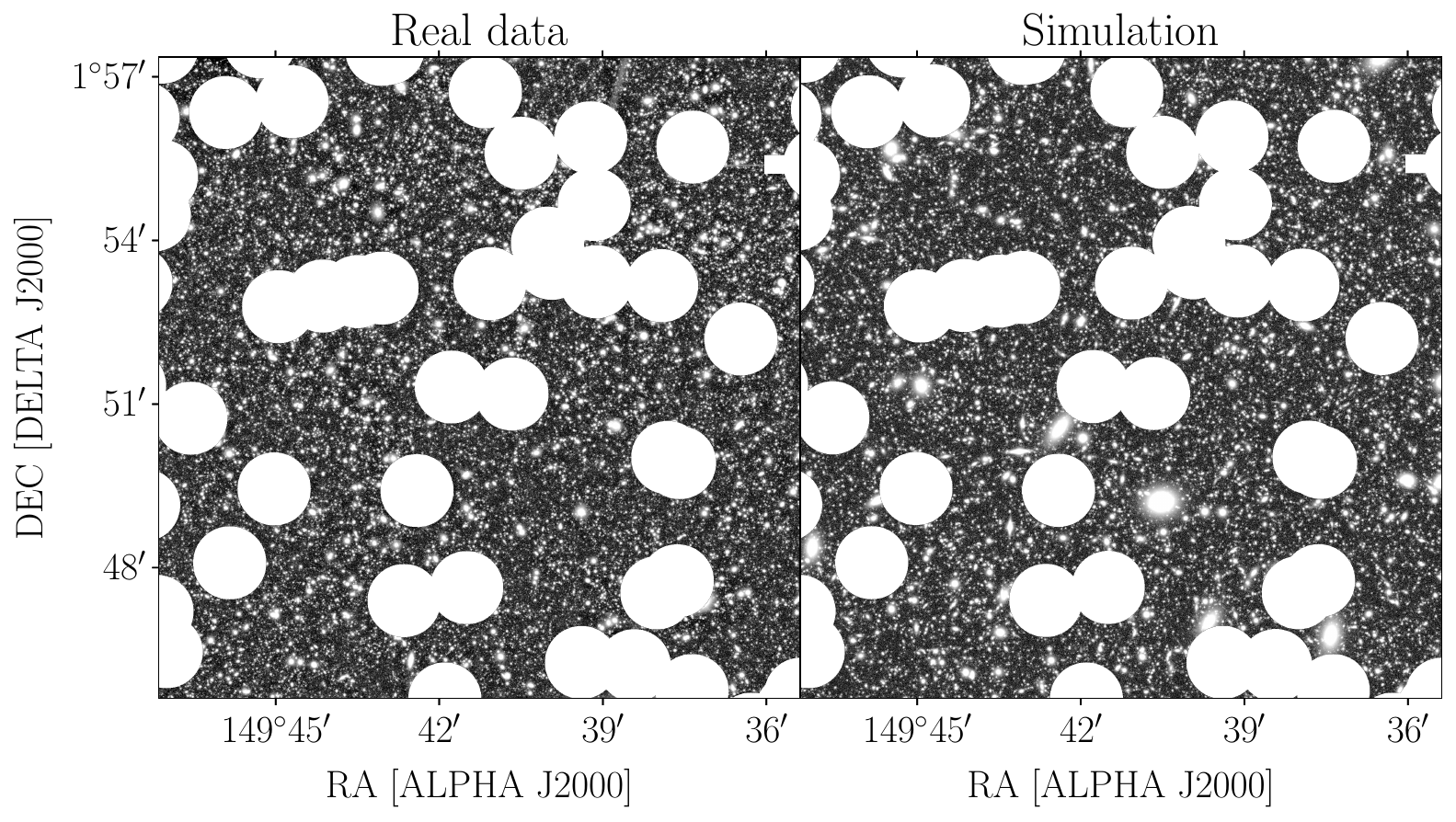}
    \caption{A comparison between a real and a simulated coadd for tract 9813, patch 702 in the HSC COSMOS field is shown. The bright star mask, as derived from the data, is applied to both the real data and simulation.}
    \label{fig:real_vs_sim}
\end{figure}

The final steps of the simulation process convert photons to ADUs by dividing out the effective gain and saturate pixels that are above the maximum value of the real data. This is a simplistic estimate of the saturation limit, which is good enough in practice since the saturated areas are always masked. We show an example of a simulated image compared to real data in Figure \ref{fig:real_vs_sim}. The most noticeable difference between the real image and the simulation is the presence of some large galaxies in the simulation. These are not ruled out by our distance measures and need further investigation. The lack of local background subtraction in the simulated image is noticeable around bright objects.  

\subsection{Source extraction and matching}
\label{sec:sextraction}
In order to proceed in our analysis, we run \SExtractor\ in dual-image mode with the same settings (reported in appendix \ref{sec:appendix_sextractor}) on real images and simulations. We use the $i$ band image as detection image. In the simulations, the \SExtractor\ detections are matched by position and magnitude to the true properties of the injected galaxies. This procedure can have an impact on the resulting photometric properties of galaxies, on the ABC posterior and on the redshift distribution since not all \SExtractor\ detections are matched to an injected galaxy or matched correctly. We use the segmentation map produced by \SExtractor\ and find, for each detection, the overlapping simulated object that minimizes the sum of the differences $\Delta \mathrm{mag}$ between \texttt{MAG\_AUTO} and true magnitude in all bands 
\begin{equation}
    \Delta\mathrm{mag} = {\sum_{b \in g,r,i,z,y} |\mathrm{mag_b}-\texttt{MAG\_AUTO\_b}|}.
\end{equation}

This matching procedure improves on our previous technique of  matching each detected object to the closest detected object below a predefined magnitude difference, especially in the case of a crowded field. Nevertheless, the procedure to match true and detected objects and the impact of blending will need further investigation in the future.
While in the matching procedure we use \texttt{MAG\_AUTO} (which is closest to the \ufig\ true magnitude), in the following we always use \texttt{MAG\_APER} in a $3''$ aperture (referred to as \texttt{MAG\_APER3} from here on, with related quantities \texttt{MAGERR\_APER3}, \texttt{FLUX\_APER3} and \texttt{FLUXERR\_APER3}), unless otherwise specified. This induces photometric biases for bright large objects (which are larger than the fixed $3''$ aperture) but provides more reliable colours and reduces photometric biases for faint objects. \texttt{FLUX\_AUTO} is the sum of the pixel values assigned to the object and thus depends on the adaptive determination of the object's size. By injecting the same catalog in a deep and ultradeep image, we observed a dependence of \texttt{MAG\_AUTO} on exposure time, since more pixels of an object rise above the background. A selection in \texttt{MAG\_AUTO} is undesirable in our case, since we calibrate the model on deep images and then extrapolate it to ultradeep images.
\subsection{COSMOS2020 redshift assignment to HSC deep fields}
\label{sec:zsample}
Before describing our ABC scheme, we show how we incorporate redshift information from COSMOS2020 in the \SExtractor\ catalogs obtained from other deep fields. We apply the reweighting technique described in  Section 4.2 of \cite{Hoyle_2018}. We start from the validation sample introduced in Section \ref{sec:validation} where we have galaxy photometry from our own \SExtractor\ run in the HSC COSMOS field overlapping with COSMOS2020 and \texttt{LePhare} and \textsc{EAZY} photo-$z$s derived from position matching the COSMOS2020 catalog. In order to assign a redshift estimate to a target galaxy in another deep field, we first add Gaussian noise to the COSMOS galaxies' fluxes until the noise level is equal to that of the target galaxy (the images in the COSMOS field are UltraDeep and thus less noisy than in the other fields). We discard COSMOS galaxies that have larger flux errors than the target galaxy. We then match a COSMOS2020 galaxy to the target galaxy by minimizing the flux $\chi^2$
\begin{equation}
    \chi^2 \equiv \sum_b \left(\frac{f_b -f_b^{\textrm{COSMOS}}}{\sigma_b}\right)^2
\end{equation}
where $b \in g,r,i,z,y$ and $f_b$ is the \texttt{FLUX\_APER3} of an object in band $b$  and $\sigma_b$ its \texttt{FLUXERR\_APER3}. In this way, we reweight the COSMOS2020 $n(z)$ to match the colour distribution of galaxies in the image we are considering. We verify that a COSMOS2020 galaxy is never matched more than 5 times in the same image (multiple matches only happen for very bright galaxies). In the following subsections, we present two contributions to the uncertainties of the redshift distributions from COSMOS2020, beyond the photo-$z$ errors on individual objects. These are taken into account when validating the $n(z)$ derived from our forward modelling approach against COSMOS2020 photo-$z$s.

\subsection{Factorised ABC inference}
\label{sec:abc_updates_inference}
We constrain the 46 parameters of our galaxy population model using the HSC deep data and the COSMOS2020 catalog described in Section \ref{sec:data}. This data combination constitutes a unique sample to precisely constrain our galaxy population model at high redshift, given its completeness up to high magnitudes. We perform simulation-based inference (SBI) to derive a posterior distribution of the parameters of the model, since the likelihood of the observables is unknown, but we have the ability to sample from it through simulations. Our ABC scheme is similar to the one used in  \cite{Tortorelli_2020}, and involves prior-to-posterior iterations. The base idea behind ABC is that the model posterior $p(\theta|x)$, where $x$ is the observed data and $\theta$ the parameters of the model, can be approximated by $p(\theta|\rho(x,y)<\epsilon)$, where $\rho$ is a distance metric, $y$ is the simulated data and $\epsilon$ is a threshold.

The unique property of our problem is that the dataset comprises of a large number of images, which can be considered as semi-independent.
We divide our full dataset ${\bf d}$ into $N$ smaller parts ${\bf d}=d_1, \dots, d_N$.
This way we can factorize the posterior on the full dataset into posterior from its parts

\begin{equation}
    p(\theta | {\bf d}) \sim \prod_{i{=}1}^{N} p(\theta | d_i).
\end{equation}
Then we use the posterior of one part of the data as prior for another part
\begin{align}
    p(\theta | d_{1}) &\sim p(d_{1} | \theta) p(\theta), \\
    p(\theta | d_{i+1}) &\sim p(d_{i+1} | \theta) p(\theta | d_i),
\end{align}
where $p(\theta)$ is the prior on the model parameters $\theta$.
This factorization allows for efficient application of the simple rejection ABC algorithm, which allows for very low complexity of our high performance computing implementation.
In practice, we begin by sampling 10000 points from the model's prior $p(\theta)$ and using each model parameter configuration to simulate a part of the data $d_1$. We accept the parameters $\theta$ where the combined distance metric computed from the simulations falls in the 20th percentile. We then resample the obtained distribution and iterate the procedure. The details of the factorized ABC are discussed at the end of this section and in Appendix~\ref{sec:appendix_abc_run}.

We modify the ABC inference engine compared to \citep{Herbel_2017,Tortorelli_2020,Kacprzak_2020} as follows:
(i) updated distance metrics, 
(ii) modifications to the ABC iteration engine and posterior modelling,
(iii) modification of model's priors.

\subsubsection{Distance metrics}
\label{sec:abc_distances}
We use the Maximum Mean Discrepancy (MMD), a kernel two-sample test for high dimensional probability distributions \cite{Gretton_2012}, as our primary distance measure
\begin{equation}
    d_{\rm{MMD}} = \frac{1}{N(N-1)} \sum_{i,j} k(x_i, x_j) + k(y_i,y_j) - k(x_i,y_j) - k(y_i,x_j)
\end{equation}
where $x_i$ is a property of the $i$th object in the real data and $y_j$ is a property of the $j$th object in the simulated data. The kernel we use is Gaussian \begin{equation}
    k(x_i,y_j) = \exp\left(-\frac{||x_i-y_j||^2}{2\sigma}\right)
\end{equation}
with free parameter $\sigma$. We describe how we choose the value of $\sigma$ in Appendix~\ref{sec:appendix_abc_run}. We extend the input vector of the MMD compared to previous work to include 
magnitudes \texttt{MAG\_APER3}, 
sizes \texttt{FLUX\_RADIUS}, the two photo-$z$ estimates assigned as described in Section \ref{sec:zsample}, 
ellipticities calculated from windowed moments \texttt{**\_WIN\_IMAGE},
and a new variable called the \emph{flux fraction} $f_{b}$, calculated as:
\begin{equation}
f_{b} = \frac{\texttt{FLUX\_APER3\_b}}{\textstyle \sum_j \texttt{FLUX\_APER3\_j}}
\end{equation}
where \texttt{FLUX\_APER\_b} is the \SExtractor\ flux in band $b$ in a circular $3''$ aperture.
Flux fractions capture similar information as colours.
This information is technically also present in the magnitudes, but these mostly impact constraints on the luminosity function parameters.
We found that the addition of the flux fractions improved our capacity to constrain the $\bar{\alpha}_i$ parameters of the SED template coefficients distribution. Since our main goal is redshift calibration, we decide to mostly focus on colours, magnitudes and redshifts. For this reason, we
include \texttt{MAG\_APER3} and $f_b$ in all 5 bands, but only include the ellipticity and \texttt{FLUX\_RADIUS} in the reference $i$ band. Our model does not account for colour gradients in the galaxy size, so it is a reasonable approximation to only constrain the size model in the reference band. The addition of information about the galaxy profile, for instance the concentration index of galaxies (the ratio between the semi-major axes containing 50\%
and 90\% of the elliptical Petrosian flux, see e.g. \cite{Shimasaku_2001,Yamauchi_2005}), in the MMD distance could help to constrain the size model. In our experiments, we found little correlation between the observed concentration indices and morphological parameters from our model, probably due to the faintness of the selected galaxy sample, in which shapes are dominated by the point spread function.

The MMD input vector is 14-dimensional. Before calculating the MMD distance, each column is scaled, so that its mean is close to zero and its standard deviation to 1. The scaling is obtained from the real data and used throughout the analysis for both real and simulated data. Since the fraction of outliers in the COSMOS2020 catalog for galaxies above magnitude 25 is extremely large ($\simeq25\%$), we only select objects with \texttt{MAG\_APER3} < 25 in the $i$ band. The galaxy sample is dominated by faint galaxies. In order to increase the weight assigned to galaxies that are the target of current wide-field surveys, we include an MMD distance with magnitude cut \texttt{MAG\_APER3\_i}=23. This choice ensures that the photometric properties of the bright end of the galaxy population are well reproduced by our ABC posterior. Omitting this distance worsens the fit slightly. Other selection cuts to remove stars and objects with bad measurements are described in Appendix \ref{sec:appendix_abc_run}. 

 As the MMD does not capture the differences in number counts, which has an important impact on the normalization of the luminosity function, we combine it with a fractional difference in number of objects \begin{equation}
    d_{ng} = \frac{|N_{\rm{SIM}}-N_{\rm{HSC}}|}{N_{\rm{HSC}}}
\end{equation}
where $N_{\rm{SIM}}$ is the number of objects in the simulation and $N_{\rm{HSC}}$ in the real data. This distance is also included for both magnitude cuts at \texttt{MAG\_APER3\_i} $<23$ and \texttt{MAG\_APER3\_i} $<25$.

 The distance metrics in an iteration $n$ are computed for all patches that are included in that iteration ($|d_n|$). We aggregate each distance using the median, which is robust to outliers, so that we have four distances ($d_{ng,25},d_{ng,23}, d_{\rm{MMD},25}$ and $d_{\rm{MMD},23}$) for each parameter configuration. We then rescale the distribution of each distance across ABC points so that it has minimum equal to zero and median equal to 1 and finally add the distances with weights: 
\begin{equation}
    d_{\rm{comb}} = 0.1\cdot d_{ng,25} + 0.1\cdot d_{ng,23} + 0.6 \cdot d_{\rm{MMD},25} + 0.2 \cdot d_{\rm{MMD},23}.
\end{equation}
The weights are chosen to rebalance the sample and upweight bright galaxies (below $i$ magnitude of 23), which are the target of Stage III cosmological surveys, and would otherwise only account for $20\%$ of the sample. 

\subsubsection{Iteration engine and posterior modelling}
We use sets of HSC patches randomly selected without replacement and increase their number in each iteration: the first set has $|d_1|=10$ patches. Using fewer images at the beginning of the ABC allows us to eliminate very unlikely areas of parameter space without wasting computing time. We then add 1 patch at every subsequent iteration.If the sets of patches are representative of the full sample, then our factorized ABC procedure corresponds to a standard rejection ABC, with reduced computational costs. If the sub-sample of images in one iteration is biased or has local properties, then the resulting ABC posterior could be biased. We consider the probability that more than 10 randomly selected images would present a systematic bias very unlikely and aggregate the distance measures computed from the different images using a median, which is robust to outliers. This way, even if some of the patches have local effects, we still obtain reliable ABC posteriors at each iteration.

In each iteration, we simulate the $|d_n|$ patches for 10000~parameter samples and keep the 2000~samples with the lowest combined distance as the posterior.
Then, we perform a density estimation for the posterior using a Gaussian Mixture Model (GMM) with 20~components.
We draw the new 10000~samples from this GMM to create a resampled posterior, and pass it as a prior to the next iteration.
The GMM fitting is performed in a transformed space to make it easier for the GMM to capture non-Gaussian distributions. We check that the GMM fitting accurately resamples the posterior and monitor the evolution of the distance measures at each iteration. When there is no more improvement in any of the distance metrics, we stop iterating the algorithm.  This stopping condition is similar to that of \cite{Tortorelli_2020}, where we look at the evolution of each distance separately because the combined distance is rescaled differently at each iteration. We ran 23 iterations of the algorithm.
The details of these iterations are summarised in Table~\ref{tab:abc_iterations} in Appendix~\ref{sec:appendix_abc_run}. Appendix~\ref{sec:appendix_abc_model} introduces the prior that we used for the ABC run, gives an overview of all model parameters and reports the resulting mean and standard deviation of each parameter in the posterior.

\section{Results}
\label{sec:results}
In this section, we present the results obtained from tuning our galaxy population model. We iteratively performed ABC inference on randomly selected batches of images taken from the HSC deep fields and complemented with reweighted COSMOS2020 many-band photometric redshifts as described in sections \ref{sec:zsample}, \ref{sec:abc_updates_inference} and appendix \ref{sec:appendix_abc_run}. We show the resulting posterior distribution of the model parameters in the following section. Then, we use samples from the posterior to run simulations in the COSMOS field and compare the photometric properties obtained by running \SExtractor\ with the same settings on simulations and real data. We choose to use these patches as validation set, since we have redshift information for individual objects from the photo-z codes \texttt{LePhare} and \textsc{EAZY}. We conclude with a comparison of the obtained redshift distributions with the COSMOS2020 catalog at different magnitude cuts.

\begin{figure}[h]
    \centering
    \includegraphics[width=1\linewidth]{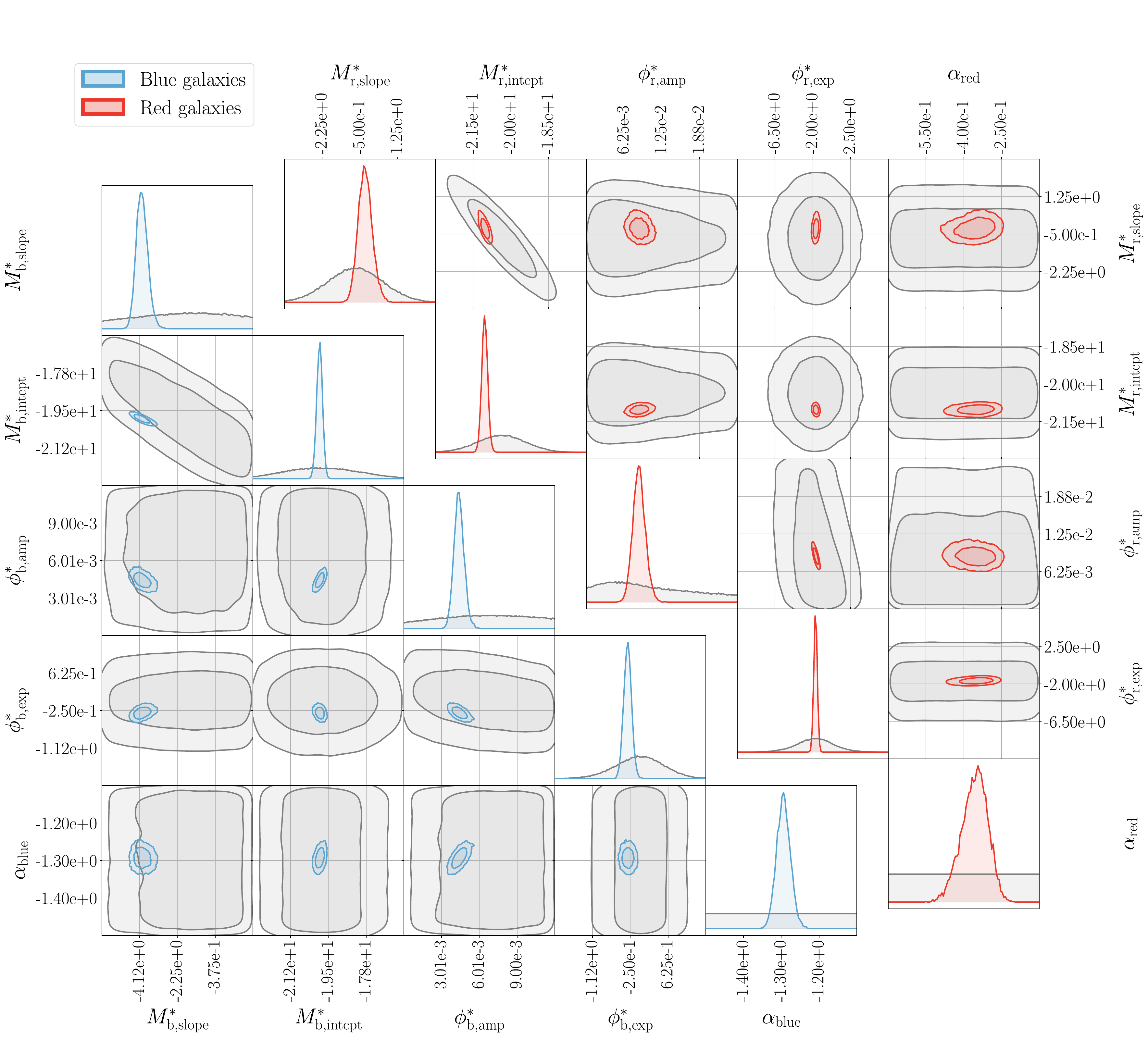}
    \caption{ABC posterior for the luminosity function parameters. The parameters for blue and red galaxies are shown in light blue and red, respectively. The ABC prior is shown in grey. }
    \label{fig:abc_posterior_lumfun}
\end{figure}

\subsection{ABC posterior}

\begin{figure}[h]
    \centering
    \includegraphics[width=1\linewidth]{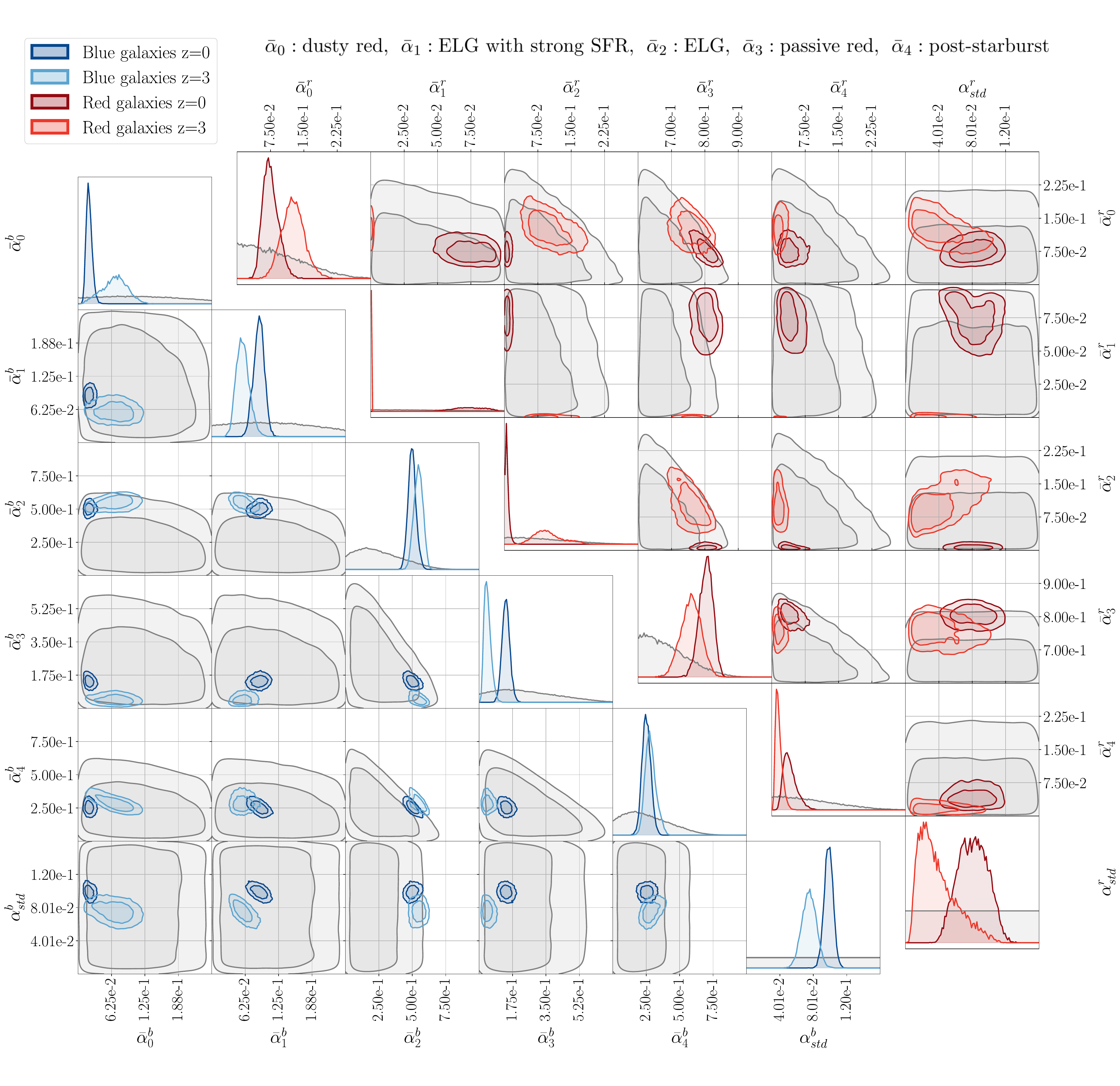}    \caption{ABC posterior for the parameters controlling the spectral energy distributions. The parameters for blue galaxies at redshift 0 are displayed in light blue and in dark blue at redshift 3. Similarly, the parameters controlling red galaxies at redshift 0 are in red and at redshift 3 in dark red. The modes of the Dirichlet coefficients are encoded in the first five parameters and one final parameter controls the standard deviation of the Dirichlet distribution. The ABC prior is shown in grey.}
    \label{fig:abc_posterior_coeffs}
\end{figure}

\begin{figure}[h]
    \centering
    \includegraphics[width=1\linewidth]{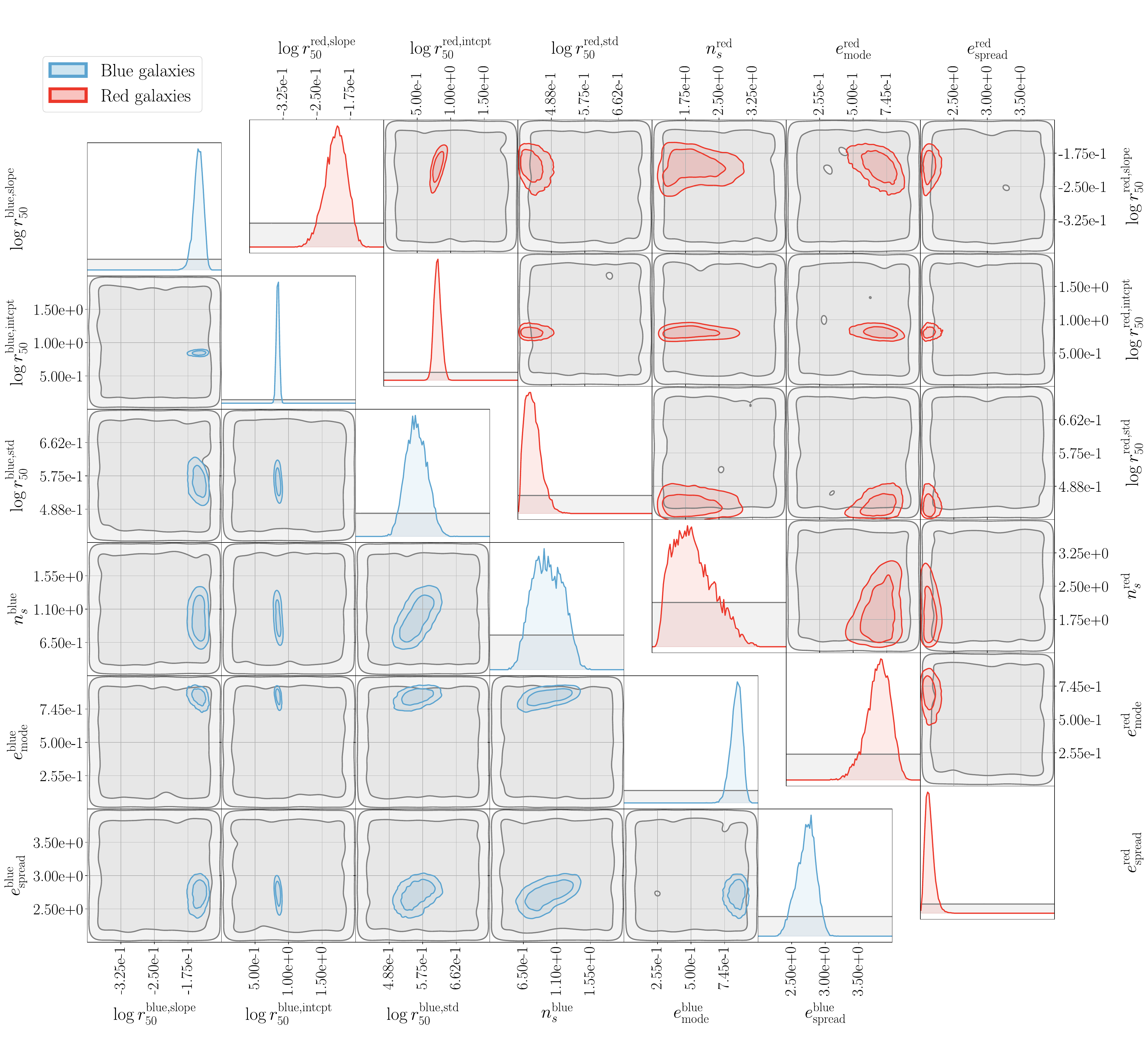}
    \caption{ABC posterior for the parameters controlling galaxy morphology. In light blue we show the parameters for blue galaxies and in red for red galaxies. The ABC prior is shown in grey.}
    \label{fig:abc_posterior_morph}
\end{figure}

Figures~\ref{fig:abc_posterior_lumfun},~\ref{fig:abc_posterior_coeffs}, and \ref{fig:abc_posterior_morph} show the posterior distribution obtained after 23 iterations of the ABC algorithm, where we fulfill our stopping condition (see Section~\ref{sec:abc_updates_inference}).
The model has 46 parameters, that we show divided into three categories for clarity: 
luminosity function parameters,
SED template coefficients and
galaxy morphology parameters. 

The luminosity function parameters are the most relevant for determining the redshift distribution. In Figure~\ref{fig:abc_posterior_lumfun} we show the parameters for blue galaxies in the lower left triangle and for red galaxies in the upper right triangle. We also plot the prior distribution in grey. We notice that the parameters for blue galaxies are better constrained than those for red galaxies, possibly because blue galaxies are more abundant. The parameters $M^*_{\mathrm{intcpt}}$ are generally very well constrained by the ABC, and we notice several strong correlations between the luminosity function parameters, most notably between $M^*_{\mathrm{intcpt}}$ and $M^*_{\mathrm{slope}}$ and between $\phi^*_{\mathrm{amp}}$ and $\phi^*_{\mathrm{exp}}$.The parameter $\alpha_{\mathrm{blue}}$ we obtain is very close to the fiducial value of $-1.3$ whereas $\alpha_{\mathrm{red}} \approx -0.35$ is slightly higher than the fiducial $-0.5$. 

In Figure~\ref{fig:abc_posterior_coeffs} we show the constraints on the SED templates coefficients. We show the prior, obtained from the catalog level ABC run described in Appendix~\ref{sec:abc_preprocess}, in grey. The posterior distribution of the parameters controlling the blue galaxy population is shown in the lower triangle of Figure~\ref{fig:abc_posterior_coeffs} (in dark blue at redshift z=0 and light blue at redshift z=3) and that controlling red galaxies in the upper triangle (in dark red for redshift z=0 and light red for redshift z=3). The $\bar{\alpha}_i$ parameters are non-negative, since a template cannot contribute less than zero to the galaxy SED. The fact that the posterior of some coefficients is very close to the lower bound of the prior is thus a sign of that coefficient not contributing to the SED of a galaxy type at redshift 0 or 3. We notice that the red galaxies are less constrained, especially at high redshift. The passive galaxy template ($T_3$) is dominating the SED of red galaxies, as imposed by the prior and there is a significant contribution of the dusty red template ($T_0$), more prominent at $z=3$. The contribution of the ELG template ($T_2$) increases with redshift, whereas that of the ELG with strong star formation template ($T_1$) decreases with redshift. This contamination indicates that our galaxy population coming from the red luminosity function is not completely passive (probably includes galaxies from the green valley). The post-startburst template ($T_4$) contributes little and decreases with redshift. The blue galaxies are better constrained and dominated by the ELG template ($T_2$). The post-starburst template ($T_4$) is also present in the blue galaxy population at all redshifts, whereas the contribution of the ELG template with strong star formation ($T_1$) slightly decreases with redshift. The contribution of the dusty red template ($T_0$) increases with redshift, whereas that of the passive red template ($T_3$) decreases with redshift.  We should not overinterpret the mixture of modes of the template coefficients, since our model allows large freedom due to the scatter parameters and the mixture of templates with redshift evolving coefficients. The \kcorrect\ templates (shown in Figure~\ref{fig:kcorrect}) are derived from SDSS data \cite{Blanton_2007} and do not provide an accurate characterization of galaxies at high redshifts. We also need to consider that there is a smooth transition between red and blue galaxies and that these two categories might not provide a good description of high redshift galaxies.

In Figure~\ref{fig:abc_posterior_morph}, we show the parameters describing galaxy morphology. These are considerably more difficult to constrain for our distance metrics, since the effects of the parameters are often degenerate and describe the full intensity profile of the galaxies, which is not sufficiently captured by ellipticity and radius alone. We again show the prior in grey and the posterior distributions for blue and red galaxies in red and blue respectively. We note that the best constrained parameter is $\log r_{50}^\mathrm{intcpt}$ for both blue and red galaxies, whereas other parameters are less constrained. The posterior on $e^{\mathrm{red}}_\mathrm{spread}$ is close to the prior boundary. This is probably due to a prior volume effect: the Beta distribution becomes uniform when $e^{\mathrm{red}}_\mathrm{spread}=2$ thus making $e^{\mathrm{red}}_\mathrm{mode}$ irrelevant. We do not extend the prior below $e^{\mathrm{red}}_\mathrm{spread}=2$, since this would correspond to U-shaped distributions of the intrinsic ellipticities which are unphysical. The posterior on $\log{r_{50}^{\mathrm{red,std}}}$ is also close to the prior boundary. We plan to extend the prior in future work. We test the impact on mean redshift of changing this parameter outside the prior range and obtain a negligible shift. The distribution of Sersic indices for red galaxies remains very broad, encompassing values between 1.5 and 3, whereas the distribution for blue Sersic indices is centered at $\approx 1$. Sersic indices lack redshift evolution in our model, which is observationally measured (high redshift galaxies have smaller Sersic indices \cite{Conselice_2014}).

\subsection{Comparison of simulations and real data}
\label{sec:photo_comparisons}
\begin{figure}[h]
    \centering
    \includegraphics[width=1\linewidth]{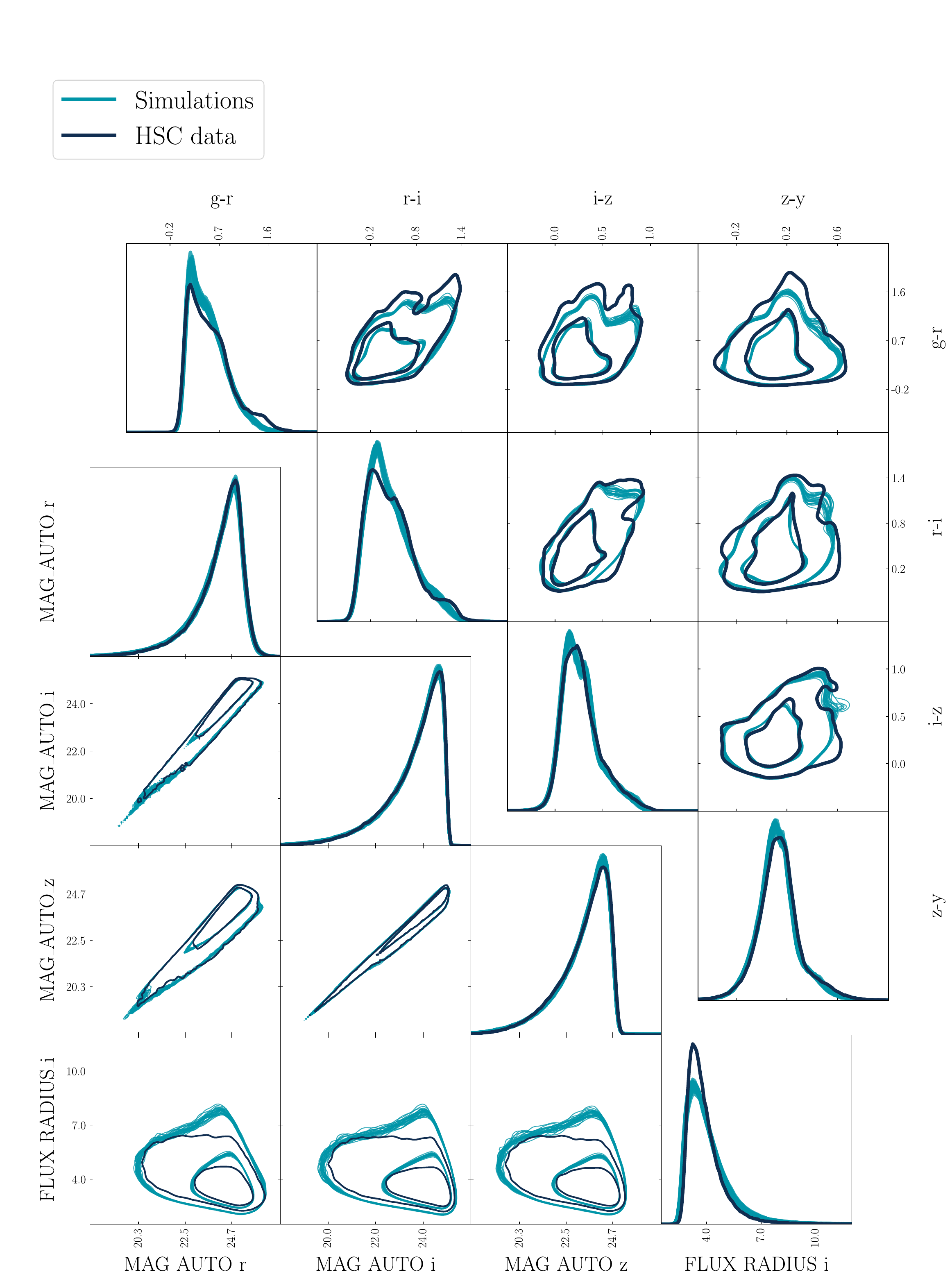}
    \caption{Comparison of selected photometric properties (\texttt{MAG\_AUTO} in $r$, $i$ and $z$ band, \texttt{FLUX\_RADIUS} in the $i$ band and colours) of HSC real data (in dark blue) and 30 simulations (in teal) in the COSMOS field.}
    \label{fig:photo_compare}
\end{figure}

We sample 30 parameter configurations from the ABC posterior at random and simulate the 56 HSC images overlapping with the COSMOS2020 COMBINED footprint. We run \SExtractor\  consistently in dual-image mode using the $i$ band image for detection on the 30 simulations and the real data and compare the obtained photometric properties of galaxies. We show 2D contours and 1D histograms of selected photometric properties of simulated galaxy samples and HSC real data up to \texttt{MAG\_APER3\_i} of 25 in Figure~\ref{fig:photo_compare}. In the lower triangle we show \texttt{MAG\_AUTO} in the $r,i,z$ bands and \texttt{FLUX\_RADIUS} in the $i$ band. The magnitudes show excellent agreement both in the 1D projections and in the 2D contours. The sizes, on the other hand, are more discrepant: there is a tail of large galaxies in the simulations and also a population of galaxies smaller than the smallest galaxies in the data. 
This is an indication of limitations in the modelling of galaxy morphology. Our model does not include size evolution with redshift at fixed absolute magnitude, which can cause a model bias in the size distribution. We do not expect the lack of size evolution with redshift to impact the redshift distributions, since our sample selection in size is very broad. We plan to explore extensions of the size model in future work. The upper triangle of Figure~\ref{fig:photo_compare} shows the colours of galaxies and their correlations. Colours are very important, since they strongly correlate with redshift. We observe a rather good agreement of the colour distributions between the simulations and the real data, with some differences in the tails of the distributions. 

\begin{figure}[t]
     \centering
     \begin{subfigure}[b]{\textwidth}
         \centering
         \includegraphics[width=0.65\textwidth]{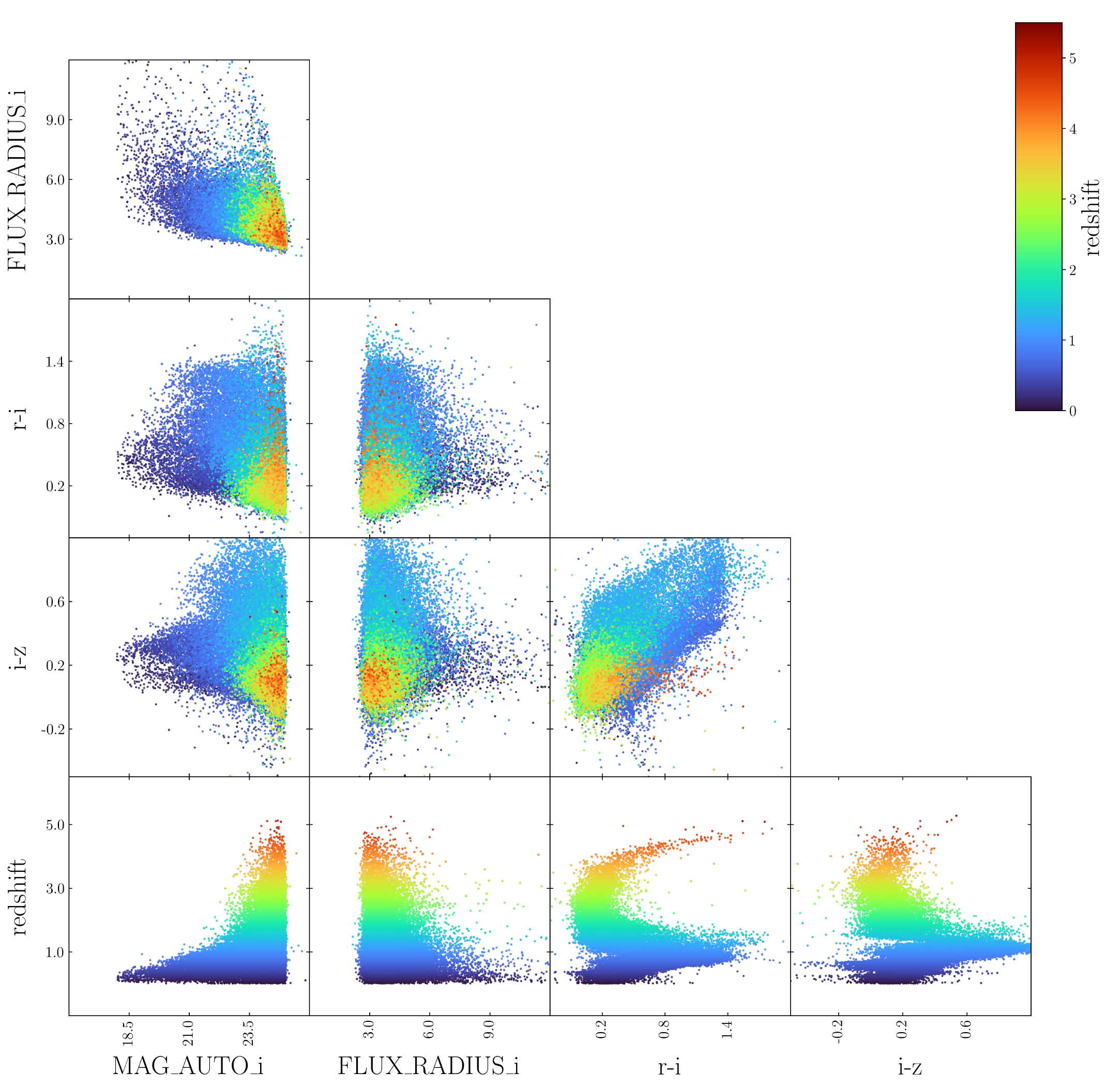}
         \caption{Scatter plot of selected photometric properties of the data in the COSMOS field. Redshift refers to \texttt{LePhare} photo-$z$s. }
     \end{subfigure}
     \hfill
     \begin{subfigure}[b]{\textwidth}
         \centering
         \includegraphics[width=0.65\textwidth]{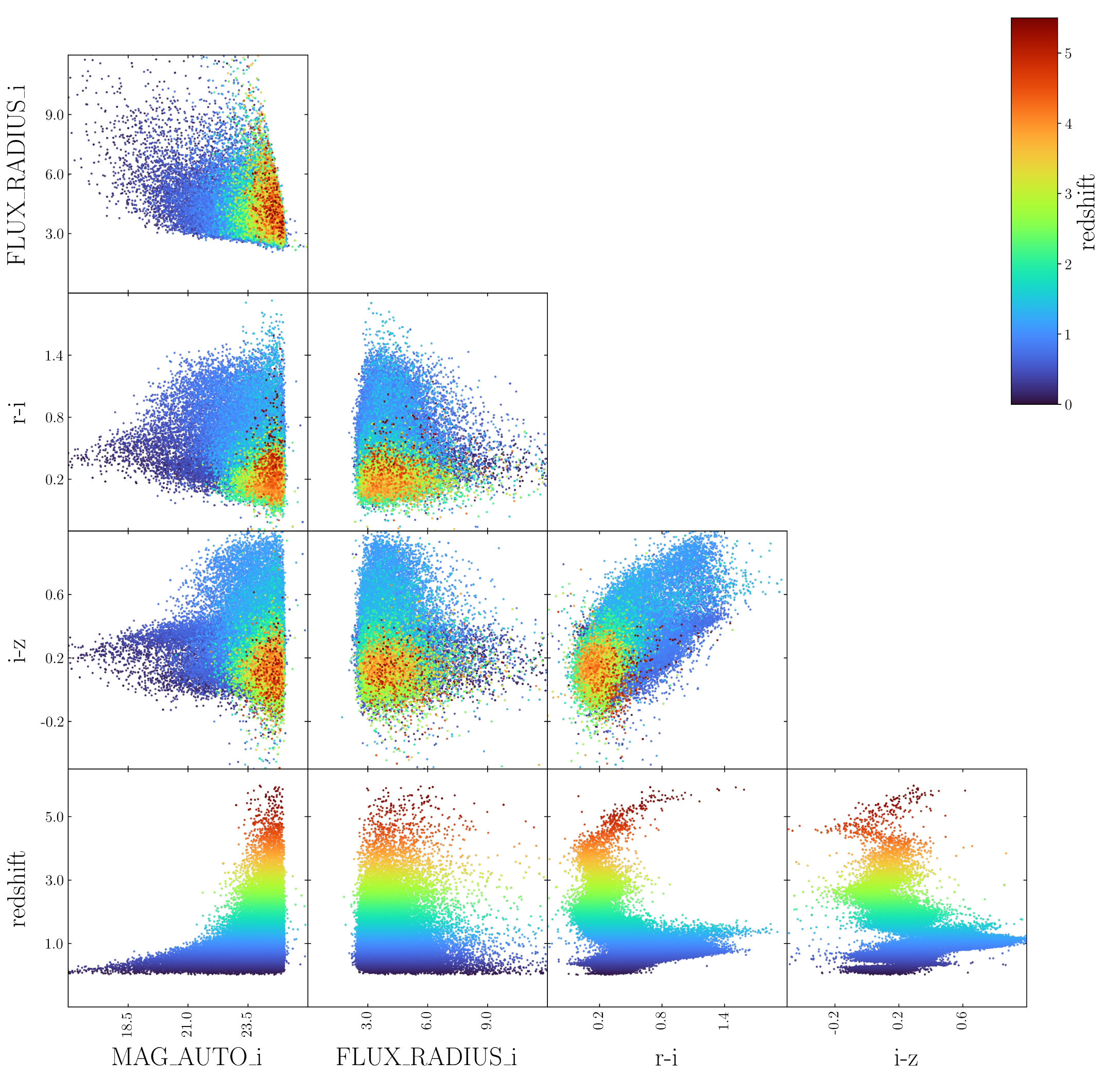}
         \caption{Scatter plot of selected photometric properties of one simulation in the COSMOS field.}
     \end{subfigure}
\caption{Scatter plots of real data and one of the simulations in the COSMOS field. We include \texttt{MAG\_AUTO} and \texttt{FLUX\_RADIUS} in the $i$ band, $r-i$ and $i-z$ colours and redshift (from COSMOS2020 CLASSIC \texttt{LePhare} in the case of the real data). Each point corresponds to a randomly selected galaxy from the catalog.}
\label{fig:scatter}
\end{figure}

In order to highlight the agreement and the discrepancies between colours, sizes and magnitudes, and also display their evolution with redshift, we include a scatter plot in Figure~\ref{fig:scatter}. We show the relation between $i$ band \texttt{MAG\_AUTO} and \texttt{FLUX\_RADIUS}, $r-i$ and $i-z$ colours and redshift (from the \texttt{LePhare} photo-$z$ code for the real data). We observe that many trends are present both in the data and simulations: in particular the colour-redshift degeneracies are well reproduced by the simulations up to $z\approx 4$. 
Since simulating realistic galaxy colours is generally a challenging task, this highlights the effectiveness and precision of this method.
We note that the simulations include an excess of galaxies at very high redshifts ($z>4.5$) that are not observed in the data. Due to the relation between absolute magnitude and size (and the evolution of angular diameter distances with redshift), these objects are large enough to be detected.  This heavy high-z tail is likely to bias the redshift distribution. The current parametrization of the luminosity function implies that the characteristic absolute magnitude $M^*(z)$ becomes brighter at higher redshifts (with a $\log(1+z)$ redshift dependence). In a Universe that grows hierarchically, this is not the case at high enough redshifts. One possibility would be to truncate the growth of $M^*(z)$ at a redshift $z_{max}$ so that $M^*(z)= \mathrm{const}$ for $z>z_{max}$ but constraining this parameter would be difficult since it would affect only a small fraction of the objects. 
We also observe an abundance of large objects in the simulations at all redshifts (as already observed in the 1D \texttt{FLUX\_RADIUS} histogram and in the comparison between real and simulated images). Our model does not account for the evolution of galaxy sizes with redshift at fixed absolute magnitude, which is seen in observations where high redshift galaxies are up to five times smaller than local galaxies (see \cite{Conselice_2014} and references therein). We leave high redshift model refinements for future work.  

\FloatBarrier

\begin{figure}[h]
     \centering
     \begin{subfigure}[]{\textwidth}
         \centering
         \includegraphics[width=\textwidth]{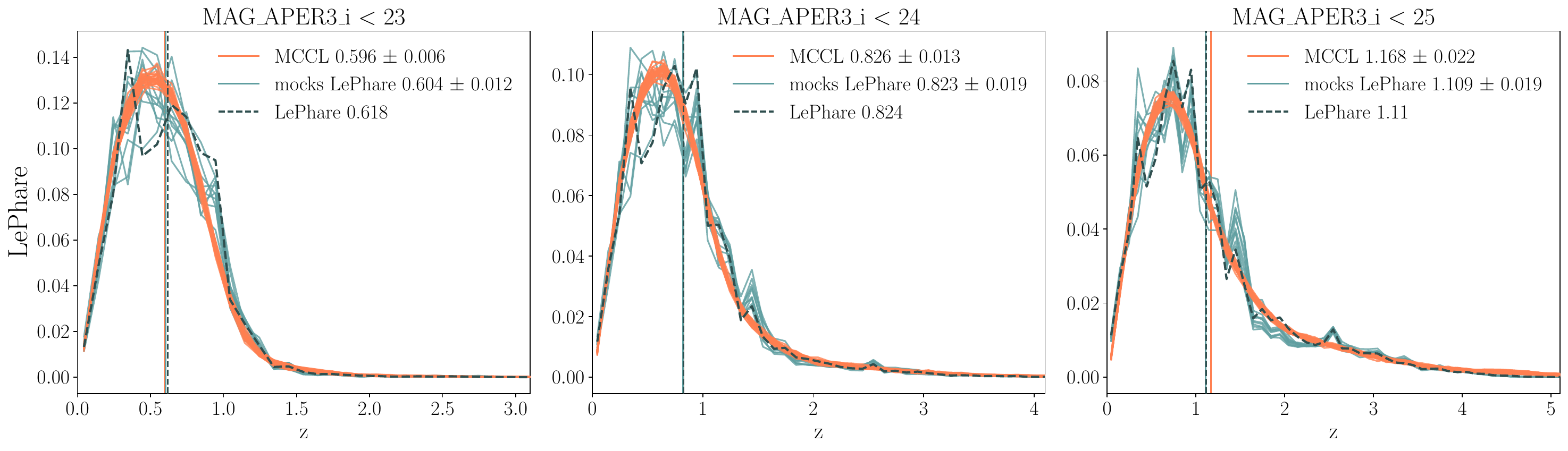}
     \end{subfigure}
     \hfill
     \begin{subfigure}[]{\textwidth}
         \centering
         \includegraphics[width=\textwidth]{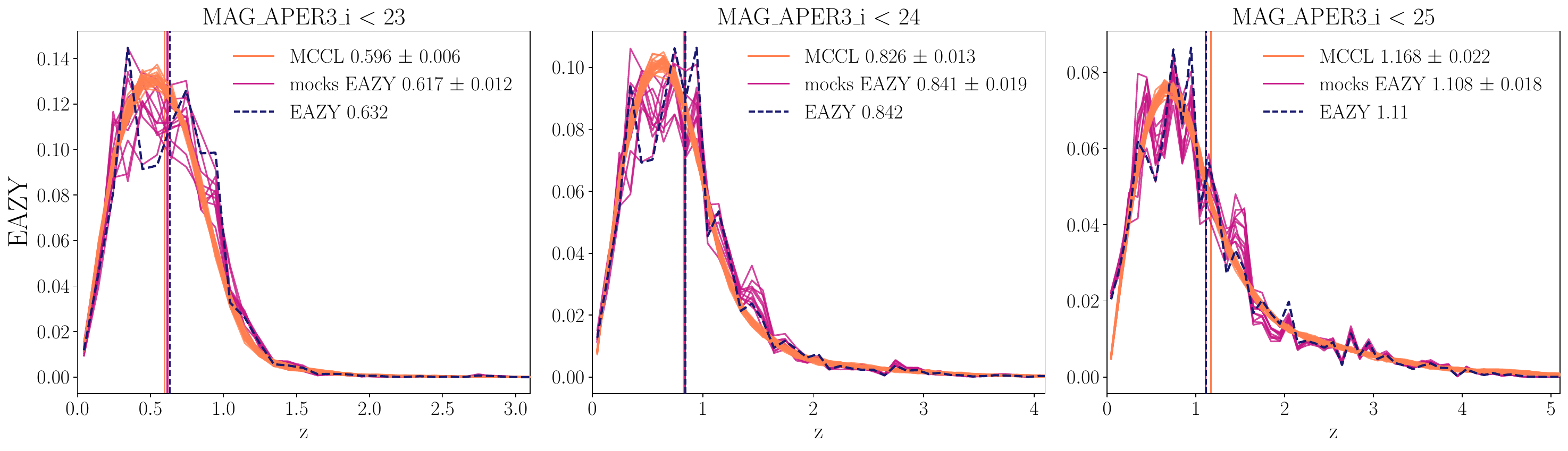}
     \end{subfigure}
\caption{Redshift distributions in the COSMOS field, from 30 simulations from the ABC posterior (in orange), from COSMOS2020 photo-$z$s and from the COSMOS mocks  generated by reweighting the COSMOS2020 photo-$z$s based on HSC deep fields photometry (as explained in Section~\ref{sec:zsample}) in 10 sets of 56 patches. The top row shows \texttt{LePhare} photo-$z$s (in dark grey for COSMOS and sea green for the mocks) and the bottom row \textsc{EAZY} photo-$z$s (in dark blue for COSMOS and purple for the mocks).}
\label{fig:nz_magcuts}
\end{figure}

\subsection{Redshift distributions}
Since the photometric properties of the simulations are in statistical agreement with the data, we derive the posterior redshift distributions. In Figure~\ref{fig:nz_magcuts} we show the $n(z)$s from 30 simulations in the COSMOS field together with those from the COSMOS2020 photometric redshift catalog in the same field. We also include the redshift distributions obtained by assigning COSMOS2020 photo-$z$s to 10 sets of 56 continuous patches selected at random in the HSC deep fields, using the reweighting procedure presented in \ref{sec:zsample} to reduce the impact of sample variance. We show three different magnitude cuts in the $i$ band \texttt{MAG\_APER3} in the three columns (\texttt{MAG\_APER3\_i} < 23, 24, 25) and the two different photo-$z$ codes used in COSMOS2020 (\texttt{LePhare} and \textsc{EAZY}) in the two rows. We notice that the redshift distributions obtained from MCCL are smooth due to the absence of clustering in the simulations.

We report mean redshift errors for our simulations, corresponding to the standard deviation of the means of the 30 simulations. We estimate an error on the mean redshift of 0.002 at all magnitude cuts from COSMOS2020 using Bootstrap. Considering sample variance and the systematic offsets between the two photo-$z$ estimates in COSMOS2020 reported in Section~\ref{sec:cosmos_sample_variance} and \ref{sec:eazy_vs_lephare}, we expect a minimum error of 0.02 in the mean redshift from COSMOS2020. By summing this in quadrature with the MCCL errors, we obtain a rough estimation of the combined errors $\sigma_{23}\approx 0.02$, $\sigma_{24}\approx 0.025$ and $\sigma_{24}\approx 0.03$ for the three magnitude cuts at \texttt{MAG\_APER3\_i}$=23,24,25$. We report the mean redshifts and errors in Table~\ref{tab:z_estimates}. Our estimated redshift distribution for $i$ band magnitude cut of 23 has low mean and is more concentrated around the mean than the $n(z)$ of COSMOS2020, especially when compared to \textsc{EAZY} photo-$z$s. This can be partly explained by the fact that the simulations are only affected by shot noise due to the limited number of objects, but do not include sample variance, since the objects are randomly distributed in space without accounting for clustering. When considering the $n(z)$ distributions reweighted according to deep field photometry in Figure~\ref{fig:nz_magcuts}, we observe the shift of $\approx 0.015$ towards lower redshift reported in Table~\ref{tab:zshifts}. To make sure that the systematic offset is not due to the reweighting methodology, we also create mocks from the simulations by assigning the redshifts from a simulated COSMOS field to equal area sets of patches from other simulated HSC deep fields. The obtained redshift shifts are negligible ($\Delta z \approx 0.001$ at all magnitude cuts). The sample variance reduction procedure leads to a 1$\sigma$ agreement between the redshift distribution of the simulations and the COSMOS2020 data at magnitude cut of 23 in the $i$ band. The redshift distribution for objects below $i$ band magnitude of 24 is in excellent agreement with COSMOS2020. The presence of heavier high redshift tails in the simulations, originating from the extrapolated redshift growth of $M^*(z)$ in the luminosity function parametrization, as explained in Section~\ref{sec:photo_comparisons}, reduces the mean redshift agreement for the \texttt{MAG\_APER3\_i} $<25$ sample to 2$\sigma$. Removing all simulated objects at $z>4.5$ reduces the MCCL mean redshift estimate of this sample to $\bar{z}_{\mathrm{MCCL},25} = 1.131 \pm 0.014$, in  better agreement with COSMOS2020.

\begin{table}[]
    \centering
    \begin{tabular}{|c|c|c|c|}
    \hline
         & \texttt{MAG\_APER3\_i} < 23 &  \texttt{MAG\_APER3\_i} < 24 & \texttt{MAG\_APER3\_i} < 25\\
         \hline
         $\bar{z}_{\mathrm{MCCL}}$& 0.596& 0.826&1.168 \\
         $\bar{z}_{\mathrm{LePhare}}$& 0.618& 0.824& 1.110\\
         $\bar{z}_{\mathrm{LePhare,\ mocks}}$& 0.604& 0.823&1.109 \\
         $\bar{z}_{\mathrm{EAZY}}$&0.632 &0.842 & 1.110\\
         $\bar{z}_{\mathrm{EAZY,\ mocks}}$&0.617 &0.841 &1.108 \\
         \hline
         $\sigma_{\mathrm{MCCL}}$ & 0.006&0.013 &0.022 \\
         $\sigma_{\mathrm{combined}}$&0.02 &0.025 & 0.03\\
         \hline
    \end{tabular}
    \caption{We report the mean redshifts obtained from MCCL in the COSMOS field, together with the \texttt{LePhare} and EAZY mean redshifts from COSMOS2020 at three different magnitude cuts (\texttt{MAG\_APER3\_i} < 23, 24, 25). We also report the \texttt{LePhare} and EAZY mean redshifts obtained when reweighting according to HSC deep photometry 10 sets of 56 images (mocks), the MCCL errors and a rough estimation of the combined errors.}
    \label{tab:z_estimates}
\end{table}

\section{Conclusions}
\label{sec:conclusion}
Redshift calibration is one of the key systematics affecting cosmic shear measurements. Shifts in the mean of the $n(z)$ lead to biased cosmological constraints from large-scale structure surveys \cite{Huterer_2006,Cunha_2012,Huterer_2013,Joudaki_2016, Hoyle_2018,Salvato_2019,Joudaki_2020, Fischbacher_2022}. It is therefore important to explore and combine a wide range of different methodologies to infer accurate photometric redshift distributions.

In this work, we presented a simulation-based inference approach to obtain redshift distributions from coadded telescope images, extending the result from \cite{Herbel_2017} to deep HSC data and increasing the accuracy of the method. This choice of dataset enables us to test our method in the regime of Stage IV surveys. We developed several extensions of the methodology, both in terms of modelling and inference. We calibrated the parameters of our galaxy population model using photometric properties from galaxies in the HSC deep fields and accurate photometric redshifts from COSMOS2020, and obtained realistic simulations. We report the resulting parameters of the model in Table~\ref{tab:abc_model}. We compared our results with photometric properties and photo-$z$s from the COSMOS2020 catalog, simulating the same area with the Subaru telescope in five broad bands, and found good agreement. We showed how sample variance in COSMOS has a strong impact on bright magnitude limited samples. We found a systematic redshift offset in the COSMOS field for objects below magnitude 23 in the HSC $i$ band, common to both photometric redshift methods (\texttt{LePhare} and \textsc{EAZY}). Previous work \cite{Joudaki_2020, Myles_2021} found that the use of the COSMOS field high quality photo-$z$s alone for redshift calibration could bias low the mean retrieved redshift. This is not in contrast with our results, since redshift calibration strongly depends on the selection function and we are only estimating the sample variance in the COSMOS field itself, and not assessing the impact of using COSMOS2020 as a calibration sample within a methodology. Once this effect is taken into account, our simulations achieve 1$\sigma$ agreement with the mean of the redshift distribution of COSMOS2020 up to \texttt{MAG\_APER3\_i}=24, and 2$\sigma$ agreement up to \texttt{MAG\_APER3\_i}=25. The overall shape of the $n(z)$ agrees well. The presence of a high redshift tail at $z>4.5$ requires further investigation and is an indication of model bias in the luminosity function.

Forward modelling has several advantages that can benefit cosmological large-scale structure surveys in different ways. On one hand, realistic simulations can be used to optimize the survey strategies and model the selection function to the needed level of accuracy. On the other hand, as done in this work, simulations can be used for calibration and to study effects that are difficult to model otherwise, for example the impact of blending and how unrecognized blends can affect the shear measurement. The MCCL method has good prospects of applicability to data from upcoming Stage IV surveys such as the Legacy Survey of Space and Time (LSST) \cite{Ivezic_2019} and Euclid \cite{Laureijs_2011,Scaramella_2022}, which will have depths comparable to the HSC deep fields. The error on the mean redshift per tomographic bin required by these surveys is $\Delta z <   0.001(1+z)$, about an order of magnitude tighter than the current work. In order to make this possible, a number of extensions and improvements are desirable. First of all, imaging of wider deep fields with many band photometry in order to reduce sample variance or deeper spectroscopy with clean selection cuts would greatly benefit photometric redshift calibration in general and forward modelling methods in particular. Secondly, it will be necessary to investigate the evolution of the luminosity function at high redshifts and find a suitable parametrization in order to avoid an excess of high redshift galaxies. This also includes designing good distance metrics to constrain the tail of the distribution using simulation-based inference. It would be beneficial to extend our galaxy population model to include effects that are well understood but not currently modelled, for example the size evolution of galaxies with redshift at fixed absolute magnitude and a relation between the absolute magnitude and colour of red galaxies (more massive galaxies are redder \cite{Baldry_2004}). 
Another possible improvement of the model of morphologies is the inclusion of bulges and disks instead of a single Sersic profile, as well as the redshift evolution of the Sersic index. A very promising path to a more physically motivated modelling is the use of stellar population synthesis (SPS) models instead of spectral templates to model the galaxies' SEDs. This entails sampling a stellar mass function rather than two luminosity functions and constructing SEDs directly from physical properties of galaxies (such as star formation rates, metallicities and gas properties). This has become feasible in terms of computing time through the emulation of SPS models \cite{Alsing_2020,Alsing_2022,Hearin_2023,Kwon_2023}. Emulators can also be used to speed up the \ufig\ simulations, by mapping the transfer function between catalogs obtained from the galaxy population model and realistic detections. This requires a more detailed understanding of selection effects and the impact of blending. Faster simulations would allow us to test extensions of the model more extensively. Finally, we have discussed how sample variance can impact redshift distributions, when the area considered is limited. In order to obtain realistic sample variance in our simulations, we need to distribute galaxies following the underlying large-scale structure. This can be achieved with the required computational speed by using Subhalo Abundance Matching (SHAM) and approximate simulations as described in \cite{Berner_2023}. 

\section*{Acknowledgement}
We thank Daniel Gr\"un, John Weaver, Will Hartley, Pascale Berner and Jamie McCullough for useful discussions. This project was supported in part by grant 200021\_192243 from the Swiss National Science Foundation. This paper is based on data collected at the Subaru Telescope
and retrieved from the HSC data archive system, which is operated
by Subaru Telescope and Astronomy Data Centre (ADC) at NAOJ. 

COSMOS2020 is based on observations collected at the European Southern Observatory under
ESO programme ID 179.A-2005 and on data products produced by CALET and
the Cambridge Astronomy Survey Unit on behalf of the UltraVISTA consortium. 

This work has made use of data from the European Space Agency (ESA) mission
{\it Gaia} (\url{https://www.cosmos.esa.int/gaia}), processed by the {\it Gaia}
Data Processing and Analysis Consortium (DPAC,
\url{https://www.cosmos.esa.int/web/gaia/dpac/consortium}). Funding for the DPAC
has been provided by national institutions, in particular the institutions
participating in the {\it Gaia} Multilateral Agreement. 

We thank the High Performance Computing group at ETH Zurich for support with the Euler cluster, which was heavily used throughout this work.  We submitted jobarrays to the cluster using \texttt{esub-epipe} \cite{Zurcher_2021,Zurcher_2023}. 
We used functionalities from several Python packages: \texttt{numpy} \cite{Harris_2020}, \texttt{scipy} \cite{Virtanen_2020}, and \texttt{matplotlib} \cite{Hunter:2007}. Corner plots were created
with \texttt{trianglechain} \cite{Kacprzak_2022,Fischbacher_2022}.
\bibliography{bibliography}
\appendix

\section{Galaxy population model priors}
\label{sec:appendix_abc_model}

We use a similar model as \cite{Herbel_2017,Refregier_2014,Tortorelli_2020}, with a number of modifications, described in Section~\ref{sec:method}.
We summarize the model with a description of model parameters, prior distributions and allowed ranges in Table~\ref{tab:abc_model}. The model has 46 parameters, but 4 of them are redundant: modes of the template coefficients $\bar{\alpha}_i$ are always forced to sum to $\sum \bar{\alpha}_i=1$.  The luminosity function parameters use the same prior as \cite{Tortorelli_2020}, with standard deviation scaled by a factor of $\times3$.
The prior column in Table~\ref{tab:abc_model} shows the distribution (Normal, Uniform, or Dirichlet) of the prior, as well as the additional bounds applied. For all variables using the Uniform distribution, a joint Sobol sequence was used to generate the prior. A suitable prior for the template coefficients $\bar{\alpha}_i$ is obtained through a catalog level ABC using the COSMOS2015 catalog \cite{Laigle_2016}. We describe this procedure in the next subsection. Table~\ref{tab:abc_model} also lists the mean and standard deviation of each parameter's 1D posterior distribution.
\afterpage{%
\begin{table*}[!ht]
\medmuskip=0mu
\setlength{\tabcolsep}{5pt}
\scriptsize
\centering
    \begin{tabular}{ p{0.03\linewidth} | p{0.12\linewidth} | p{0.41\linewidth} | p{0.16\linewidth} |
    p{0.13\linewidth}  |}
        & Parameter & Meaning & Prior & Posterior\\

        \midrule
        \multirow{10}{*}[-14mm]{\begin{sideways}Luminosity functions (10)\end{sideways}} 

        & $M^*_\mathrm{b,slope}$ 
        & Slope of the redshift evolution of the parameter $M^*$ in the Schechter LF for blue galaxies, see Equation~\ref{eq:m_star_evol} 
        & Prior from \cite{Tortorelli_2020} $\times 3$, $\in~[-6, 1.5]$ & $-4.0 \pm 0.3$
        \\  \cmidrule{2-5}

        & $M^*_\mathrm{b,intcpt}$ 
        & Intercept of the redshift evolution of the parameter $M^*$ in the Schechter LF for blue galaxies, see Equation~\ref{eq:m_star_evol}
        & Prior from \cite{Tortorelli_2020} $\times 3$, $\in~[-23,-16]$ & $-19.9 \pm 0.1$ 
        \\ \cmidrule{2-5}

        & $M^*_\mathrm{r,slope}$ 
        & Slope of the redshift evolution of the parameter $M^*$ in the Schechter LF for red galaxies, see Equation~\ref{eq:m_star_evol} 
        &  Prior from \cite{Tortorelli_2020} $\times 3$,\quad $\in~[-4,3]$ & $-0.3 \pm 0.3$
        \\ \cmidrule{2-5}

        & $M^*_\mathrm{r,intcpt}$ 
        & Intercept of the redshift evolution of the parameter $M^*$ in the Schechter LF for red galaxies, see Equation~\ref{eq:m_star_evol} 
        &  \mbox{Prior from \cite{Tortorelli_2020} $\times 3$},\quad $\in~[-23,-17]$ & $-21.0\pm 0.1$
        \\ \cmidrule{2-5}

        & $\phi^*_\mathrm{b,exp}$ 
        & Exponent of the redshift evolution of the parameter $\phi^*$ in the Schechter LF for blue galaxies, see Equation~\ref{eq:phi_star_evol} 
        & Prior from \cite{Tortorelli_2020} $\times 3$,\quad $\in~[-2,1.5]$ & $-0.31 \pm 0.09$
        \\ \cmidrule{2-5}

        & $\phi^*_\mathrm{b,amp}$ 
        & Amplitude of the redshift evolution of the parameter $\phi^*$ in the Schechter LF for blue galaxies, see Equation~\ref{eq:phi_star_evol} 
        & Prior from \cite{Tortorelli_2020} $\times 3$,\quad $\in~[\num{1.1}{-5}, \num{1.2}{-2}]$ & $0.0044\pm0.0004$ 
        \\ \cmidrule{2-5}

        & $\phi^*_\mathrm{r,exp}$ 
        & Exponent of the redshift evolution of the parameter $\phi^*$ in the Schechter LF for red galaxies, see Equation~\ref{eq:phi_star_evol} 
        & Prior from \cite{Tortorelli_2020} $\times 3$,\quad $\in~[-11,7]$ & $-1.7\pm0.2$ 
        \\ \cmidrule{2-5}

        & $\phi^*_\mathrm{r,amp}$ 
        & Amplitude of the redshift evolution of the parameter $\phi^*$ in the Schechter LF for red galaxies, see Equation~\ref{eq:phi_star_evol} 
        & Prior from \cite{Tortorelli_2020} $\times 3$, $\in~[\num{2}{-8}, \num{2.5}{-2}]$ & $0.009\pm0.001$
        \\ \cmidrule{2-5}

        & $\alpha_\mathrm{blue}$ 
        & Steepness of the faint-end slope in the Schechter LF for blue galaxies, see Equation~\ref{eq:schechter}
        & $\mathcal{U}[-1.5,-1.1]$ & $-1.29\pm0.02$
        \\ \cmidrule{2-5}

        & $\alpha_\mathrm{red}$ 
        & Steepness of the faint-end slope in the Schechter LF for red galaxies, see Equation~\ref{eq:schechter} 
        & $\mathcal{U}[-0.7,-0.1]$ & $-0.36\pm0.05$ 
        \\ 
        
        \midrule
        \multirow{6}{*}[-6mm]{\begin{sideways}Galaxy morphology (12) \end{sideways}} 

        & $\log{r_{50}}^\mathrm{blue/red}_\mathrm{slope}$ 
        & Slope of the evolution of the average intrinsic physical size of galaxies with absolute magnitude 
        &  $\mathcal{U}[-0.4,-0.1]$ & \mbox{b: $-0.15\pm0.01$} \mbox{r: $-0.21\pm0.02$}
        \\ \cmidrule{2-5}

        & $\log{r_{50}}^\mathrm{blue/red}_\mathrm{intcpt}$ 
        & Intercept of the evolution of the average intrinsic physical size of galaxies with absolute magnitude 
        & $\mathcal{U}[0,2]$ 
        & \mbox{b: $0.84\pm0.02$} \mbox{r: $0.81\pm0.05$}
        \\ \cmidrule{2-5}

        & $\log{r_{50}}^\mathrm{blue/red}_\mathrm{std}$ 
        & Standard deviation of the normal distribution we use to sample intrinsic physical galaxy sizes 
        &  $\mathcal{U}[0.4, 0.75]$ 
        & \mbox{b: $0.56\pm0.03$} \mbox{r: $0.44\pm0.02$}
        \\ \cmidrule{2-5}

        & $n_s^\mathrm{blue}$ 
        & Mode of the Sersic index distribution of blue galaxies 
        & $\mathcal{U}[0.2, 2]$ & $1.0\pm0.2$
        \\ \cmidrule{2-5}

        &  $n_s^\mathrm{red}$ 
        & Mode of the Sersic index distribution of red galaxies 
        & $\mathcal{U}[1, 4]$ & $2.0\pm0.4$
        \\ \cmidrule{2-5}
        
        & $e_\mathrm{mode}^{\rm{blue/red}}$ 
        & Ellipticity distribution mode for blue/red galaxies 
        & $\mathcal{U}[0.01, 0.99]$ & \mbox{b: $0.83\pm0.04$} \mbox{r: $0.69\pm0.08$}
        \\ \cmidrule{2-5}

        & $e_\mathrm{spread}^{\rm{blue/red}}$ 
        & Ellipticity distribution spread for blue/red galaxies 
        & $\mathcal{U}[2, 4]$ & \mbox{b: $2.7\pm0.1$} \mbox{r: $2.14\pm0.06$}
        \\ 
        
        \midrule
        \multirow{2}{*}{\begin{sideways}\shortstack{SED coeff. (24)}\end{sideways}}

        & $\bar{\alpha}^{\rm{blue/red}}_{i, 0/3}$ & Normalized Dirichlet concentration parameters at $z{=}0/3$ from which the template coefficients for blue/red galaxies are sampled, $i{=}0,\dots,4$, $\sum_i \bar{\alpha}_i{=}1$ 
        & 5-dimensional Dirichlet, see \ref{sec:abc_preprocess} for details & See Table~\ref{tab:sed_params_post}
        \\ \cmidrule{2-5}

        & $\alpha^{\rm{blue/red}}_{ \mathrm{std}, 0/3}$ 
        & Standard deviation of the normalized Dirichlet concentration parameters at $z{=}0/3$ from which the template coefficients for blue/red galaxies are sampled 
        & $\mathcal{U}[1e-4, 0.16]$ & See Table~\ref{tab:sed_params_post}
        \\ 
        \midrule
   \end{tabular}

\caption{
Table with galaxy population model parameters, priors and resulting 1D posteriors. Luminosity function is shortened as LF. The details of the prior ranges of the Dirichlet distributions for the template coefficients are explained in subsection~\ref{sec:abc_preprocess}. We briefly describe the model parameters and report the type of prior and the its boundaries.
The last column shows the resulting mean value and standard deviation from the ABC posterior.
}
\label{tab:abc_model}
\end{table*}
\clearpage}
\subsection{Template coefficient priors from COSMOS2015 catalog}
\label{sec:abc_preprocess}
As mentioned in Section \ref{sec:abc_updates_model}, we do not rely on the weights derived in \cite{Herbel_2017}  using the New York University
Value-Added Galaxy Catalog to differentiate the SED between red and blue galaxies, but impose different priors on the Dirichlet $\bar{\alpha}_i$ parameters for the two galaxy types. 
This is motivated by the changes to the spectral energy distribution modelling described in Section~\ref{sec:abc_updates_model}, that make the model more interpretable.
To capture redshift-colour dependencies of higher redshift galaxies, we derive these priors from a comparison with the COSMOS2015 catalog \cite{Laigle_2016}.

First, we select galaxies from this catalog using the following cuts: 
$z\in[0.3, 4]$, $\mathrm{mag}_{\mathrm{ip}} \in [10,24.5]$, $\texttt{TYPE}==0$.
The comparison between the simulated and observed galaxies is performed using their redshift and colours, defined with respect to the reference band.
We use colour and redshift to avoid constraining the luminosity function, and exploit just the colour-redshift information.
We use the reference band $\mathrm{ip}$ and compute the colours as a difference with bands $\mathrm{NUV, u, B, V, r, zpp, Y, J, H, Ks}$. 
For the simulated galaxies, these magnitudes are calculated using \ufig\ up to catalog generation, with the use of filters provided in the COSMOS2015 dataset.
The comparison is performed for red and blue galaxies separately. In the real data we use the \texttt{CLASS} provided by COSMOS2015 to separate between star-forming and quiescent galaxies (classified using NUV--r /r -- J diagram).
We use the nearest-neighbour estimator of Universal Divergence \citep{Wang2009unidiv} as a distance metric between simulations and real data.

We do not intend to create a posterior on the template coefficient values, but rather find the upper and lower limits on the Dirichlet modes of template coefficient values $\bar{\alpha}_i$.
We then perform an iterative procedure of progressively narrowing down the ranges for all coefficients.
Starting with a uniform range for $\bar{\alpha}_i \in [0,1]$, we generate 10000 samples from the luminosity function prior described in Section~\ref{sec:abc_updates_model}.
We then calculate Universal Divergence between the redshifts and colours from the simulated and COSMOS galaxies.
The columns are scaled before the comparison, using a robust scaler from the \texttt{scikit-learn} package.
We select 2000 best points and calculate the lower and upper limits on $\bar{\alpha}_i$.
We then input these new limits and generate another 10000 samples with them.
We repeat this process 20 times.
The ranges for the coefficients obtained from this procedure define the prior in the main ABC run and are shown in Table~\ref{tab:alpha_priors}, with coefficients rounded roughly to 0.1. The obtained ranges agree with expectations for both blue and red galaxies, with $\bar{\alpha}_2$ and $\bar{\alpha}_3$ dominating for the blue and red galaxies, respectively. We report the means and standard deviations of the obtained posterior of the SED template coefficients in Table~\ref{tab:sed_params_post}.

\begin{table}[h]
    \centering
    \begin{tabular}{|c|c|c|c|c|c|c|c|c|c|c|}
        \hline
        parameter & $\bar{\alpha}_{0,z}^{\mathrm{blue}}$ & $\bar{\alpha}_{1,z}^{\mathrm{blue}}$ & $\bar{\alpha}_{2,z}^{\mathrm{blue}}$ & $\bar{\alpha}_{3,z}^{\mathrm{blue}}$ & $\bar{\alpha}_{4,z}^{\mathrm{blue}}$& $\bar{\alpha}_{0,z}^{\mathrm{red}}$ & $\bar{\alpha}_{1,z}^{\mathrm{red}}$ & $\bar{\alpha}_{2,z}^{\mathrm{red}}$ & $\bar{\alpha}_{3,z}^{\mathrm{red}}$ & $\bar{\alpha}_{4,z}^{\mathrm{red}}$ \\
        \hline
        lower limit & 0 & 0 & 0 & 0 & 0 & 0 & 0 & 0 & 0.6 & 0 \\

        upper limit & 0.25 & 0.25 & 1 & 0.7 & 1 & 0.3 & 0.1 & 0.3 & 1 & 0.3 \\
        \hline
    \end{tabular}
    \caption{Upper and lower limits on the modes of the Dirichlet coefficients $\bar{\alpha}_i$ for blue and red galaxies derived from the catalog level ABC. The same boundaries are imposed at redshifts $z=0$ and $z=3$. These are the upper and lower limits of the Dirichlet priors of the subsequent ABC run on HSC DUD data. }
    \label{tab:alpha_priors}
\end{table}

    \begin{table}[h]
    \centering
    \setlength{\tabcolsep}{3.5pt}
    \begin{tabular}{
    |c|c|c|c|c|c|c|}
    \hline
      type, $z$ & $\bar{\alpha_0}$ & $\bar{\alpha_1}$ & $\bar{\alpha_2}$ & $\bar{\alpha_3}$ & $\bar{\alpha_4}$ & $\alpha_{\mathrm{std}}$ \\
       \hline
       blue, $z=0$ & $0.021\pm0.005$& $0.09\pm 0.01$& $0.50\pm 0.03$& $0.14 \pm 0.02$& $0.25\pm0.03$ & $0.099 \pm 0.005$ \\
       blue, $z=3$ & $0.07\pm0.02$ & $0.06\pm0.01$ & $0.55\pm0.03$ & $0.04\pm0.02$ & $0.28\pm0.04$ & $0.075\pm0.009$ \\
       red, $z=0$ & $0.08\pm0.02$ & $0.07\pm0.01$ & $0.006\pm0.004$ & $0.80\pm0.02$ & $0.04\pm0.02$ & $0.08\pm0.02$ \\
       red, $z=3$ & $0.13\pm0.02$ & $1e-5\pm2e-5$ & $0.10\pm0.03$ & $0.76\pm0.03$ & $0.016\pm0.008$ & $0.04\pm0.02$ \\
       \hline
    \end{tabular}
    \caption{ABC posterior means and standard deviations of the SED coefficients for red and blue galaxies at redshift $z=0$ and $z=3$.}
    \label{tab:sed_params_post}
\end{table}

\section{Details of the ABC runs}
\label{sec:appendix_abc_run}

In this Appendix, we describe the details of our ABC analysis that were omitted in Section \ref{sec:abc_updates_inference}.
The ABC iteration engine is similar to the one presented in \cite{Tortorelli_2020} and depends on a sequence of prior-to-posterior iterations.

\paragraph{Iterations}
We start by sampling 10000~parameter configurations from the prior defined in Table~\ref{tab:abc_model}, with limits in the Dirichlet coefficients $\bar{\alpha}_i$ from Table~\ref{tab:alpha_priors}. We discard samples when the simulation fulfills one of  the rejection criteria: (i) having more than 1 million blue or red galaxies (ii) having less than 300 or more than 20000~objects below magnitude 24 in the $i$ band. These are considered extreme conditions, that no simulation that is similar to the real data would fulfil and help us restrict to more likely parts of parameter space, without an excessive use of computing time.
For each of the 10000~configurations, we simulate 10~HSC patches in the first iteration, as described in Section~\ref{sec:abc_updates_inference}. 
In the following iterations, we increase the number of simulated patches by 1, whereas the number of parameter configurations is fixed to 10000. Table~\ref{tab:abc_iterations} shows the number of patches simulated per iteration and the corresponding sky area in $\mathrm{deg}^2$.  

\begin{table*}[b]
\medmuskip=0mu
\setlength{\tabcolsep}{5pt}
\scriptsize
\centering

\begin{tabular}{| c | c | c |}
\hline
iteration & $N_{p,\rm{sim}}$  &  sky area \\
\hline
1  & 10     & 0.38 deg$^2$\\
2  & 11     & 0.42 deg$^2$\\
3  & 12     & 0.46 deg$^2$ \\
4  & 13     & 0.5 deg$^2$\\
5  & 14     & 0.54  deg$^2$\\
6  & 15     & 0.58 deg$^2$\\
7  & 16     & 0.61 deg$^2$ \\
8  & 17     & 0.65 deg$^2$\\
9  & 18     & 0.69 deg$^2$\\
10  & 19     & 0.73 deg$^2$\\
11  & 20     & 0.77 deg$^2$\\
12  & 21     & 0.8 deg$^2$\\
13  & 22     & 0.85 deg$^2$\\
14  & 23     & 0.88 deg$^2$\\
15  & 24     & 0.92 deg$^2$\\
16  & 25     & 0.96 deg$^2$\\
17  & 26     & 1 deg$^2$\\
18 & 27 &    1.04 deg$^2$\\
19 & 28 &    1.08 deg$^2$ \\
20 & 29 &    1.11 deg$^2$ \\
21 & 30 &    1.15 deg$^2$ \\
22 & 31 &    1.19 deg$^2$\\
23 & 32 &    1.23 deg$^2$\\
\hline
\end{tabular}
\caption{
Number of patches used in each iteration and corresponding sky area covered by the HSC DUD patches.}
\label{tab:abc_iterations}
\end{table*}

\paragraph{Sample selection}
The distance measures described in Section~\ref{sec:abc_distances} are computed using the \SExtractor\ catalogs, created in all $grizy$ bands, based on the detection in the $i$ band. 
We perform the PSF estimation using a Convolutional Neural Network \cite{Herbel_2018}, in the same way as in \cite{Kacprzak_2020}.
We run \SExtractor\ on the HSC data first, and then on the simulated images during the ABC iterations.
From the catalogs, we select galaxies with the following set of cuts applied in all bands with strict \texttt{and} conditions:
\begin{align*}
\label{eqn:abc_cuts}
\small
&\texttt{FLAGS} <4, \quad
14 < \texttt{MAG\_APER3} < 30,  \quad \texttt{MAG\_AUTO} < 99, \quad
0.1 < \texttt{FLUX\_RADIUS} <100,  \nonumber \\
& -3 < \log_{10}{(\texttt{SNR})} \equiv \log_{10}{\left(\frac{\texttt{FLUX\_AUTO}}{\texttt{FLUXERR\_AUTO}}\right)} < 4, \quad 0 < \texttt{ELL} < 1, \\ & \texttt{N\_EXPOSURES} > 0, \quad 
0.5 < r_{50} / \texttt{PSF\_FWHM}, \quad \texttt{CLASS\_STAR} < 0.95\\
\end{align*}
where \texttt{ELL} is the absolute ellipticity calculated from windowed moments \texttt{**\_WIN\_IMAGE}, \texttt{N\_EXPOSURES} is the number of exposures in the coadd at the position of the object, $r_{50}$ is the object size defined as $ r_{50} = 2\cdot\ln(2)\cdot(\texttt{X2\_WIN\_IMAGE}+\texttt{Y2\_WIN\_IMAGE})^{1/2}$, as in \cite{Kacprzak_2020}.
Both the PSF size cut and the \texttt{CLASS\_STAR} cut are applied to create a pure galaxy sample. In addition, we impose $\texttt{MAG\_APER3} < 25$ in the $i$ band. Note that the \SExtractor\ detections are matched in the simulations to the true properties of the injected galaxies, as explained in Section \ref{sec:HSC-forward-model},
so that a further criterion for simulated objects to be selected for the MMD distances is that the detection has been matched to a
true simulated object. We additionally require that the objects do not lie on the image mask.
When computing $d_{\mathrm{MMD},23}$ we additionally impose \texttt{MAG\_APER3\_i} < 23.

\paragraph{Optimization of the kernel radius parameter}
To obtain the most sensitive MMD distance, it is common to optimize the parameters of the kernel used to compute it \cite{Gretton_2012}.
We use a Gaussian kernel with a single parameter $\sigma$, corresponding to the correlation scale. We compute $\sigma$ for the different MMD distances as the median distance between samples drawn from the same probability distribution (the real data) \cite{Herbel_2017, Gretton_2012}.

\paragraph{Modelling of posterior distributions}
We create the posterior distribution at each iteration by setting the 20th percentile as a threshold, thus selecting the 2000 out of 10000 samples with the smallest combined distance. 
We then create a model of this posterior using a Gaussian Mixture Model (GMM), from the \texttt{scikit-learn} implementation\footnote{\url{scikit-learn.org/stable/modules/generated/sklearn.mixture.GaussianMixture.html}}.
We use 20 Gaussians to fit the distribution.
Before fitting, we transform the model parameter samples to a gaussianized space.
This is done to make the GMM more suited for fitting the distribution, especially for parameters with uniform priors.
First we rescale the parameters to lie between $\in[\num{1}{-8}, 1-\num{1}{-8}]$, and then apply a Gaussian inverse-CDF transform.
We draw 10000~new samples in the gaussianized space using Sobol sampling and invert them back to the original space.
We verify that the GMM model in the transformed space is a good representation of the posterior by comparing the 2D~marginal projections of all parameter combinations for the original 2000~samples and 10000~new GMM samples. The GMM samples from the model posterior are passed as priors to the next iteration of the ABC algorithm.

\section{\SExtractor\ settings}
\label{sec:appendix_sextractor}
We report in Table \ref{tab:sextractor} the \SExtractor\ configuration.

\begin{table}[]
    \centering
    \begin{tabular}{c|c}
    \hline
        \textbf{\SExtractor\ parameter} & \textbf{Value} \\
        \hline
        CATALOG\_TYPE & FITS\_1.0 \\
        DETECT\_TYPE & CCD \\
        DETECT\_MINAREA & 5 \\
        THRESH\_TYPE & RELATIVE \\
        DETECT\_THRESH & 1.5 \\
        ANALYSIS\_THRESH & 1.5 \\
        FILTER & Y \\
        FILTER\_NAME & gauss\_3.0\_5x5.conv \\
        DEBLEND\_NTHRESH & 32 \\
        DEBLEND\_MINCONT & 0.00001 \\
        CLEAN & Y \\
        CLEAN\_PARAM & 1.0 \\
        MASK\_TYPE & CORRECT \\
        MAG\_ZEROPOINT & 27 \\
        PIXEL\_SCALE & 0.168 \\
        STARNNW\_NAME & default.nnw \\
        BACK\_SIZE & 128 \\
        BACK\_FILTERSIZE & 3 \\
        BACKPHOTO\_TYPE & LOCAL \\
        BACKPHOTO\_THICK & 24 \\
        WEIGHT\_TYPE & NONE \\
    \end{tabular}
    \caption{\SExtractor\ configuration used in this work both on real images and simulations. The missing parameters change per patch and band and are described in the text.}
    \label{tab:sextractor}
\end{table}

\end{document}